\documentclass[format=acmsmall, review=false, screen=true]{acmart}

\usepackage{booktabs} 

\setcopyright{rightsretained}
\acmVolume{4} 
\acmNumber{CSCW2}
\acmJournal{PACMHCI}
\acmYear{2020}
\acmMonth{10}
\copyrightyear{2020}
\acmArticle{148}

\acmDOI{10.1145/XXXXXXX}

\received{April 2019}
\received[revised]{July 2019}
\received[revised]{January 2020}
\received[revised]{April 2020}
\received[accepted]{July 2020}

\setcopyright{ccbysa}

\usepackage[ruled]{algorithm2e} 

\SetAlFnt{\small}
\SetAlCapFnt{\small}
\SetAlCapNameFnt{\small}
\SetAlCapHSkip{0pt}
\IncMargin{-\parindent}

\usepackage{balance}       
\usepackage{graphics}      
\usepackage{todonotes}     
\usepackage{color}
\usepackage{soul}
\usepackage{booktabs}
\usepackage{subcaption}
\usepackage{textcomp}
\PassOptionsToPackage{warn}{textcomp}

\usepackage{microtype}        

\newcommand\leadin[1]{%
    \vskip 5pt \noindent\textbf{#1.} %
}

\begin{document}

\title[ORES]{ORES: Lowering Barriers with Participatory Machine Learning in Wikipedia}

\newcommand\highlightcolor{black}
\author{Aaron Halfaker}
\orcid{0000-0001-8907-6367}
\affiliation{%
  \institution{Microsoft}
  \streetaddress{1 Microsoft Way}
  \city{Redmond}
  \state{WA}
  \postcode{98052}
  \country{USA}}
\email{aaron.halfaker@gmail.com}
\authornote{The majority of this work was authored when Halfaker was affiliated with the Wikimedia Foundation.}

\author{R. Stuart Geiger}
\affiliation{%
  \department{Department of Communication}
  \department{Hal\i c\i o{\u g}lu Data Science Institute}
  \institution{University of California, San Diego}
  \streetaddress{9500 Gilman Drive}
  \city{San Diego}
  \state{CA}
  \postcode{92093}
  \country{USA}}
\email{stuart@stuartgeiger.com}
\authornote{The majority of this work was authored when Geiger was affiliated with the Berkeley Institute for Data Science at the University of California, Berkeley.}

\renewcommand{\shortauthors}{Halfaker \& Geiger}

\begin{abstract}
Algorithmic systems---from rule-based bots to machine learning classifiers---have a long history of supporting the essential work of content moderation and other curation work in peer production projects.  From counter-vandalism to task routing, basic machine prediction has allowed open knowledge projects like Wikipedia to scale to the largest encyclopedia in the world, while maintaining quality and consistency.  However, conversations about how quality control should work and what role algorithms should play have generally been led by the expert engineers who have the skills and resources to develop and modify these complex algorithmic systems. In this paper, we describe ORES: an algorithmic scoring service that supports real-time scoring of wiki edits using multiple independent classifiers trained on different datasets. ORES decouples several activities that have typically all been performed by engineers: choosing or curating training data, building models to serve predictions, auditing predictions, and developing interfaces or automated agents that act on those predictions. This meta-algorithmic system was designed to open up socio-technical conversations about algorithms in Wikipedia to a broader set of participants.  In this paper, we discuss the theoretical mechanisms of social change ORES enables and detail case studies in participatory machine learning around ORES from the 5 years since its deployment.

\end{abstract}

%
%


%
%


\maketitle
\settopmatter{printfolios=true}
\section{Introduction}
\label{sec:introduction}
Wikipedia --- the free encyclopedia that anyone can edit --- faces many challenges in maintaining the quality of its articles and sustaining the volunteer community of editors. The people behind the hundreds of different language versions of Wikipedia have long relied on automation, bots, expert systems, recommender systems, human-in-the-loop assisted tools, and machine learning to help moderate and manage content at massive scales. The issues around artificial intelligence in Wikipedia are as complex as those facing other large-scale user-generated content platforms like Facebook, Twitter, or YouTube, as well as traditional corporate and governmental organizations that must make and manage decisions at scale. And like in those organizations, Wikipedia's automated classifiers are raising new and old issues about truth, power, responsibility, openness, and representation.

Yet Wikipedia's approach to AI has long been different than in corporate or governmental contexts typically discussed in emerging fields like Fairness, Accountability, and Transparency in Machine Learning (FAccTML) or Critical Algorithms Studies (CAS). The volunteer community of editors has strong ideological principles of openness, decentralization, and consensus-based decision-making. The paid staff at the non-profit Wikimedia Foundation --- which legally owns and operates the servers --- are not tasked with making editorial decisions about content\footnote{Except in rare cases, such as content that violates U.S. law, see \url{http://enwp.org/WP:OFFICE}}. Content review and moderation, in either its manual or automated form, is instead the responsibility of the volunteer editing community. A self-selected set of volunteer developers build tools, bots, and advanced technologies, generally with some consultation with the community. Even though Wikipedia's prior socio-technical systems of algorithmic governance have generally been more open, transparent, and accountable than most platforms operating at Wikipedia's scale, ORES\footnote{\url{https://ores.wikimedia.org} and \url{http://enwp.org/:mw:ORES}}, the system we present in this paper, pushes even further on the crucial issue of who is able to participate in the development and use of advanced technologies.

ORES represents several innovations in openness in machine learning, particularly in seeing openness as a socio-technical challenge that is as much about scaffolding support as it is about open-sourcing code and data~\cite{selbst_fairness_2019}. With ORES, volunteers can curate labeled training data from a variety of sources for a particular purpose, commission the production of a machine classifier based on particular approaches and parameters, and make this classifier available via an API which anyone can query to score any edit to a page --- operating in real time on the Wikimedia Foundation's servers. Currently, 110 classifiers have been produced for 44 languages, classifying edits in real-time based on criteria like ``damaging / not damaging,'' ``good-faith / bad-faith,'' language-specific article quality scales, and topic classifications like ``Biography" and ``Medicine". ORES intentionally does not seek to produce a single classifier to enforce a gold standard of quality, nor does it prescribe particular ways in which scores and classifications will be incorporated into fully automated bots, semi-automated editing interfaces, or analyses of Wikipedian activity. As we describe in section~\ref{sec:design_rationale}, ORES was built as a kind of cultural probe~\cite{hutchinson2003technology} to support an open-ended set of community efforts to re-imagine what machine learning in Wikipedia is and who it is for.

Open participation in machine learning is widely relevant to both researchers of user-generated content platforms and those working across open collaboration, social computing, machine learning, and critical algorithms studies. ORES implements several of the dominant recommendations for algorithmic system builders around transparency and community consent~\cite{crawford2016algorithm,diakopoulos2015algorithmic,sandvig2014auditing}. We discuss practical socio-technical considerations for what openness, accountability, and transparency mean in a large-scale, real-world, user-generated content platform. Wikipedia is also an excellent space for work on participatory governance of algorithms, as the broader Wikimedia community and the non-profit Wikimedia Foundation are founded on ideals of open, public participation. All of the work presented in this paper is publicly-accessible and open sourced, from the source code and training data to the community discussions about ORES. Unlike in other nominally `public' platforms where users often do not know their data is used for research purposes, Wikipedians have extensive discussions about using their archived activity for research, with established guidelines we followed\footnote{See \url{http://enwp.org/WP:NOTLAB} and \url{http://enwp.org/WP:Ethically_researching_Wikipedia}}. This project is part of a longstanding engagement with the volunteer communities which involves extensive community consultation, and the case studies research has been approved by a university IRB.

{\color{\highlightcolor}
In discussing ORES, we could present a traditional HCI ``systems paper'' and focus on the architectural details of ORES, a software system we constructed to help expand participation around ML in Wikipedia. However, this technology only advances these goals when they are used in particular ways by our\footnote{Halfaker led the development of ORES and the Platform Scoring Team while employed at the Wikimedia Foundation. Geiger was involved in some conceptualization of ORES and related projects, as well as observed the operation of ORES and the team, but was not on the team developing or operating ORES, nor affiliated with the Wikimedia Foundation.} engineering team at the Wikimedia Foundation and the volunteers who work with us on that team to design and deploy new models. As such, we have written a new genre of a \textit{socio-technical systems paper}. We have not just developed a technical system with intended affordances; ORES as a technical system is the `kernel' \cite{ribes2014kernel} of a socio-technical system we operate and maintain in certain ways. In section~\ref{sec:ores_the_technology}, we focus on the technical capacities of the ORES service, which our needs analysis made apparent. Then in section~\ref{sec:ores_the_socio-technical_system}, we expand to focus on the larger socio-technical system built around ORES. 

We detail how this system enables new kinds of activities, relationships, and governance models between people with capacities to develop new machine learning classifiers and people who want to train, use, or direct the design of those classifiers. We discuss many \emph{demonstrative cases} --- stories from our work with members of Wikipedia's various language communities in collaboratively designing training data, engineering features, building models, and expanding the technical affordances of ORES to meet new needs. These demonstrative cases provide a brief exploration of the complex, community-governed response to the deployment of novel AIs in their spaces. The cases illustrate how crowd-sourced and crowd-managed auditing can be an important community governance activity. Finally, we conclude with a discussion of the issues raised by this work beyond the case of Wikipedia and identify future directions.
}

\section{Related work}
\label{sec:related_work}
\subsection{The politics of algorithms}
Algorithmic systems \cite{seaver2017algorithms} play increasingly crucial roles in the governance of social processes \cite{gillespie2014relevance,eubanks2018automating,benjamin2019race}. Software algorithms are increasingly used in answering questions that have no single right answer and where using prior human decisions as training data can be problematic~\cite{barocas2013governing}. Algorithms designed to \emph{support work} change people's work practices, shifting how, where, and by whom work is accomplished \cite{crawford2016algorithm, zuboff1988age}. Software algorithms gain political relevance on par with other process-mediating artifacts (e.g. laws \& norms~\cite{lessig1999code}).

There are repeated calls to address power dynamics and bias through transparency and accountability of the algorithms that govern public life and access to resources~\cite{buolamwini2018gender,diakopoulos2017algorithmic,sandvig2014auditing}. The field around effective transparency, explainability, and accountability mechanisms is growing. We cannot fully address the scale of concerns in this rapidly shifting literature, but we find inspiration in Kroll et al's discussion of the limitations of auditing and transparency \cite{kroll2016accountable}, Mulligan et al's shift towards the term ``contestability'' \cite{mulligan2019shaping}, Geiger's call to go ``beyond opening up the black box'' \cite{geiger2017beyond}, and Selbst et al.'s call to explore the socio-technical/societal context and include social actors of algorithmic systems when considering issues of fairness \cite{selbst_fairness_2019}.

In this paper, we discuss a specific socio-political context --- Wikipedia's algorithmic quality control and socialization practices --- and the development of novel algorithmic systems for support of these processes.  We implement a meta-algorithmic intervention aligned with Wikipedians' principles and practices: deploying a service for building and deploying prediction algorithms, where many decisions are delegated to the volunteer community. Instead of training the single best classifier and implementing it in our own designs, we embrace having multiple potentially-contradictory classifiers, with our \emph{intended} and \emph{desired} outcome involving a public process of training, auditing, re-interpreting, appropriating, contesting, and negotiating both these models themselves and how they are deployed in content moderation interfaces and automated agents. Often, work on technical and social ways to achieve fairness and accountability does not discuss this broader kind of socio-infrastructural intervention on communities of practice, instead staying at the level of  models themselves.

However, CSCW and HCI scholarship has often addressed issues of algorithmic governance in a broader socio-technical lens, which is in line with the field's longstanding orientation to participatory design, value-sensitive design, values in design, and collaborative work~\cite{friedman1996value,friedman1996bias,schmidt1992taking,harrison2007,schuler1993}. Decision support systems and expert systems are classic cases which raise many of the same lessons as contemporary machine learning~\cite{bentley1992,star1994steps,lynch2004user}. Recent work has focused on more participatory~\cite{lee2019webuildai}, value-sensitive~\cite{zhu2018value}, and experiential design~\cite{alvarado2018towards} approaches to machine learning. CSCW and HCI has also long studied human-in-the-loop systems \cite{zhang2017,choe2014,geiger2010work}, as well as explored implications of ``algorithm-in-the-loop'' socio-technical systems~\cite{green2019principles}. Furthermore, the field has explored various adjacent issues as they play out in citizen science and crowdsourcing, which is often used as a way to collect, process, and verify training data sets at scale, to be used for a variety of purposes~\cite{wiggins2011conservation,harandi2018,kittur2013future}.

\subsection{Machine learning in support of open production}
Open peer production systems, like most user-generated content platforms,  use machine learning for content moderation and task management. For Wikipedia and related Wikimedia projects, quality control for removing ``vandalism''\footnote{''Vandalism'' is an emic term in the English-language Wikipedia (and other language versions) for blatantly damaging edits that are routinely made to articles, such as adding hate speech, gibberish, or humor; see \url{https://enwp.org/WP:VD}.} and other inappropriate edits to articles is a major goal for practitioners and researchers.  Article quality prediction models have also been explored and applied to help Wikipedians focus their work in the most beneficial places and explore coverage gaps in article content.

\leadin{Quality control and vandalism detection} The damage detection problem in Wikipedia is one of great scale.  English Wikipedia is edited about 142k times each day, which immediately go live without review.  Wikipedians embrace this risk, but work tirelessly to maintain quality. Damaging, offensive, and/or fictitious edits can cause harms to readers, the articles' subjects, and the credibility of all of Wikipedia, so all edits must be reviewed as soon as possible ~\cite{geiger2010work}. As an information overload problem, filtering strategies using machine learning models have been developed to support the work of Wikipedia's patrollers (see \cite{adler2011wikipedia} for an overview). Some researchers have integrated their prediction models into purpose-designed tools for Wikipedians to use (e.g. STiki \cite{west2010stiki}, a classifier-supported moderation tool). Through these machine learning models and constant patrolling, most damaging edits are reverted within seconds of when they are saved ~\cite{geiger2013levee}.

\leadin{Task routing and recommendation}
Machine learning plays a major role in how Wikipedians decide what articles to work on. Wikipedia has many well-known content coverage biases (e.g. for a long period of time, coverage of women scientists lagged far behind~\cite{halfaker2017interpolating}). Past work has explored collaborative recommender-based task routing strategies (see SuggestBot~\cite{cosley2007suggestbot}), in which contributors are sent articles that need improvement in their areas of expertise. Such systems show strong promise to address content coverage biases, but could also inadvertently reinforce biases.

{\color{\highlightcolor}
\subsection{Community values applied to software and process}
Wikipedia has a large community of ``volunteer tool developers'' who build bots, third party tools, browser extensions, and Javascript-based gadgets, which add features and create new workflows. This ``bespoke'' code can be developed and deployed without seeking approval or major actions from those who own and operate Wikipedia's servers~\cite{geiger2014bots}.  The tools play an oversized role in structuring and governing the actions of Wikipedia editors~\cite{geiger2011lives,tkacz2014wikipedia}.  Wikipedia has long managed and regulated such software, having built formalized structures to mitigate the potential harms that may arise from automation. English Wikipedia and many other wikis have formal processes for the approval of fully-automated bots \cite{geiger2011lives}, which are effective in ensuring that robots do not often get into conflict with each other~\cite{geiger2017operationalizing}.

Some divergences between the bespoke software tools that Wikipedians use to maintain Wikipedia and their values have been less apparent.  A line of critical research has studied the unintended consequences of this complex socio-technical system, particularly on newcomer socialization~\cite{halfaker2013rise,morgan2013tea,halfaker2014snuggle}.  In summary, Wikipedians struggled with the issues of scaling when the popularity of Wikipedia grew exponentially between 2005 and 2007~\cite{halfaker2013rise}.  In response, they developed quality control processes and technologies that prioritized efficiency by using machine prediction models~\cite{halfaker2014snuggle} and templated warning messages~\cite{halfaker2013rise}.  This transformed newcomer socialization from a human and welcoming activity to one far more dismissive and impersonal~\cite{morgan2013tea}, which has caused a steady decline in Wikipedia's editing population~\cite{halfaker2013rise}.  The efficiency of quality control work and the elimination of damage/vandalism was considered extremely politically important, while the positive experience of a diverse cohort of newcomers was less so. 

After the research about these systemic issues came out, the political importance of newcomer experiences in general and underrepresented groups specifically was raised substantially.  But despite targeted efforts and shifts in perception among some members of the Wikipedia community~\cite{narayan2015effects, morgan2013tea}\footnote{See also a team dedicated to supporting newcomers \url{http://enwp.org/:m:Growth\_team}}, the often-hostile quality control processes that were designed over a decade ago remain largely unchanged~\cite{halfaker2014snuggle}.  Yet recent work by Smith et al. has confirmed that ``positive engagement" is one of major convergent values expressed by Wikipedia editors with regards to algorithmic tools. Smith et al. discuss conflicts between supporting efficient patrolling practices and maintaining positive newcomer experiences.~\cite{smith2020keeping}.  Smith et al. and Selbst et al. highlight discuss how it is crucial to examine not only how the algorithm behind a machine learning model functions, but also how predictions are surfaced to end-users and how work processes are formed around the predictions~\cite{smith2020keeping, selbst_fairness_2019}.
}

\section{Design rationale}
\label{sec:design_rationale}
{\color{\highlightcolor}
In this section, we discuss mechanisms behind Wikipedia's socio-technical problems and how we as socio-technical system builders designed ORES to have impact within Wikipedia.  Past work has demonstrated how Wikipedia's problems are systemic and caused in part by inherent biases in the quality control system. To responsibly use machine learning in addressing these problems, we examined how Wikipedia functions as a distributed system, focusing on how processes, policies, power, and software come together to make Wikipedia happen.

\subsection{Wikipedia uses decentralized software to structure work practices}
In any online community or platform, software code has major social impact and significance, enabling certain features and supporting (semi-)automated enforcement of rules and procedures; or, ``code is law" ~\cite{lessig1999code}. However, in most online communities and platforms, such software code is built directly into the server-side code base. Whoever has root access to the server has jurisdiction over this sociality of software. 

For Wikipedia's almost twenty year history, the volunteer editing community has taken a more decentralized approach to software governance. While there are many features integrated into server-side code --- some of which have been fiercely-contested  \cite{neotarf2014media}\footnote{\url{https://meta.wikimedia.org/wiki/Superprotect}} --- the community has a strong tradition of users developing third-party tools, gadgets, browser extensions, scripts, bots, and other external software. The MediaWiki software platform has a fully-featured Application Programming Interface (API) providing multiple output formats, including JSON\footnote{JSON stands for Javascript Object Notation and it is designed to be a straightforward, human-readable data format}. This has let developers write software that adds or modifies features for Wikipedians who download or enable them. 

As Geiger describes, this model of ``bespoke code" is deeply linked to Wikipedians' conceptions of governance, fostering more decentralized decision-making ~\cite{geiger2014bots}. Wikipedians develop and install third-party software without generally needing approvals from those who own and operate Wikipedia's servers. There are some centralized governance processes, especially for fully-automated bots, which give developer-operators massive editorial capacity. Yet even bot approval decisions are made through local governance mechanisms of each language version, rather than top-down decisions by Wikimedia Foundation staff.

In many other online communities and platforms, users who want to implement an addition or change to user interfaces or ML models would need to convince the owners of the servers. In contrast, the major blocker in Wikipedia is typically finding someone with the software engineering and design expertise to develop such software, as well as the resources and free time to do this work. This is where issues of equity and participation become particularly relevant: Ford \& Wajcman discuss how Wikipedia's infrastructural choices have added additional barriers to participation, where programming expertise is important in influencing encyclopedic decision-making \cite{ford2017anyone}.  Expertise, resources, and free time to overcome such barriers are not equitably distributed, especially around technology~\cite{sculley2015hidden}. This is a common pattern of inequities in ``do-ocracies," including the content of Wikipedia articles. 

The people with the skills, resources, time, and inclination to develop such software tools and ML models have held a large amount of power in deciding what types of work will and will not be supported ~\cite{niederer2010wisdom,geiger2011lives,muller2013work,tkacz2014wikipedia,livingstone2016population}. Almost all of the early third-party tools (including the first tools to use ML) were developed by and/or for so-called ``vandal fighters," who prioritized the quick and efficient removal of potentially damaging content. Tools that supported tasks like mentoring new users, edit-a-thons, editing as a classroom exercise, or identifying content gaps lagged significantly behind, even though these kinds of activities were long recognized as key strategies to help build the community, including for underrepresented groups. Most tools were also written for the English-language Wikipedia, with tools for other language versions lagging behind.

\subsection{Using machine learning to scale Wikipedia}
Despite the massive size of Wikipedia's community (66k avg. monthly active editors in English Wikipedia in 2019\footnote{\url{https://stats.wikimedia.org//en.wikipedia.org/contributing/active-editors/normal|line|2-year|~total|monthly}}) --- or precisely because of this size --- the labor of Wikipedia's curation and support processes is massive. More edits from more contributors requires more content moderation work, which has historically resulted in labor shortages. For example, if patrollers in Wikipedia review 10 revisions per minute for vandalism (an aggressive estimate, for those seeking to catch only blatant issues) it would require 483 labor hours \emph{per day} just to review the 290k edits saved to all the various language editions of Wikipedia.\footnote{\url{https://stats.wikimedia.org//all-wikipedia-projects/contributing/user-edits/normal|bar|1-year|~total|monthly}}  In some cases, the labor shortage has become so extreme that Wikipedians have chosen to shut down routes of contribution in an effort to minimize the workload.\footnote{English Wikipedia now prevents the direct creation of articles by new editors ostensibly to reduce the workload of patrollers.  See \url{https://en.wikipedia.org/wiki/Wikipedia:Autoconfirmed_article_creation_trial} for an overview.} Content moderation work can also be exhausting and even traumatic. \cite{roberts2019behind}

When labor shortages threatened the function of the open encyclopedia, ML was a breakthrough technology. A reasonably fit vandalism detection model can filter the set of edits that need to be reviewed down by 90\%, with high levels of confidence that the remaining 10\% of edits contains almost all of the most blatant kinds of vandalism -- the kind that will cause credibility problems if readers see these edits on articles.  The use of such a model turns a 483 labor hour problem into a 48.3 labor hour problem. This is the difference between needing 240 coordinated volunteers to work 2 hours per day to patrol for vandalism and needing 24 coordinated volunteers to work for 2 hours per day. For smaller wikis, it means that vandalism can be tackled by just 1 or 2 part-time volunteers, who can spend time on other tasks. Beyond vandalism, there are many other cases where sorting, routing, and filtering make problems of scale manageable in Wikipedia. 

\subsection{Machine learning resource distribution}
\label{sec:machine_learning_resource_distribution}
If these crucial and powerful third-party workflow support tools only get built if someone with expertise, resources, and time freely decides to build them, then the barriers to and resulting inequities in developing and implementing ML at Wikipedia's scale are even more stark.  Those with advanced skills in ML and data engineering can have day jobs that prevent them from investing the time necessary to maintain these systems ~\cite{sculley2015hidden}. Diversity and inclusion are also major issues in computer science education and the tech industry, including gender, race/ethnicity, and national origin \cite{margolis2002unlocking,margolis2017stuck}. Given Wikipedia's open participation model but continual issues with diversity and inclusion, it is also important to note that free time is not equitably distributed in society ~\citep{Bianchi2010}.

To the best of our knowledge, in the 15 years of Wikipedia's history prior to ORES, there were only three ML classifier projects that successfully built and served real-time predictions at the scale of at least one entire language version of a Wikimedia project. All of these also began on the English language Wikipedia, with only one ML-assisted content moderation tool \footnote{\url{http://enwp.org/WP:Huggle}} supporting other language versions. Two of these three were developed and hosted at university research labs. These projects also focused on supporting a particular user experience in an interface the model builders also designed, which did not easily support re-use of their models in other tools and interfaces.  One notable exception was the public API that Stiki's\footnote{\url{http://enwp.org/WP:STiki}} developer made available for a period of time.  While limited in functionality,\footnote{Only edits made while Stiki was online were scored so no historical analysis was possible and many gaps in time persisted.} in past work, we appropriated this API in a new interface designed to support mentoring ~\cite{halfaker2014snuggle}. We found this experience deeply inspirational about how publicly hosting models with APIs could support new and different uses of ML.

Beyond expertise and time, there are financial costs in hosting real-time ML classifier services at Wikipedia's scale. These require continuously operating servers, unlike third-party tools, extensions, and gadgets that run on a user's computer. The classifier service must keep in sync with the new edits, prepared to classify any edit or version of an article. This speed is crucial for tasks that involve reviewing new edits, articles, or editors, as ML-assisted patrollers respond within 5 seconds of an edit being made ~\cite{geiger2013levee}.  Keeping this service running at scale, at high reliability, and in sync with Wikimedia's servers is a computationally intensive task requiring costly server resources.\footnote{ORES is currently hosted in two redundant clusters of 9 high power servers with 16 cores and 64GB of RAM each. Along with supporting services, ORES requires 292 CPU cores and 1058GB of RAM.}


Given this background context, we intended to foster collaborative model development practices through a model hosting system that would support shared use of ML as a common infrastructure. We initially built a pattern of consulting with Wikipedians about how they made sense of their important concepts --- like damage, vandalism, good-faith vs. bad-faith, spam, quality, topic categorization, and whatever else they brought to us and wanted to model. We saw an opportunity to build technical infrastructure that would serve as the base of a more open and participatory socio-technical system around ML in Wikipedia. Our goals were twofold: First, we wanted to have a much wider range of people from the Wikipedian community substantially involved in the development and refinement of new ML models. Second, we wanted those trained ML models to be broadly available for appropriation and re-use for volunteer tool developers, who have the capacity to develop third-party tools and browser extensions using Wikipedia's JSON-based API, but less capacity to build and serve ML models at Wikipedia's scale.
}

\section{ORES the technology}
\label{sec:ores_the_technology}
{\color{\highlightcolor}
We built ORES as a technical system to be multi-purpose in its support of ML in Wikimedia projects. Our goal is to support as broad of use-cases as possible, including ones that we could not imagine. As Figure \ref{fig:ores_data_user} illustrates, ORES is a \textit{machine learning as a service} platform that connects sources of labeled training data, human model builders, and live data to host trained models and serve predictions to users on demand. As an endpoint, we implemented ORES as an API-based web service, where once a project-specific classifier had been trained and deployed, predictions from a classifier for any specific item on that wiki (e.g. an edit or a version of a page) could be requested via an HTTP request, with responses returned in JSON format. For Wikipedia's existing communities of bot and tool developers, this API-based endpoint is the default way of engaging with the platform.

\begin{figure*}[h!]
  \centering
  \includegraphics[width=.95\textwidth]{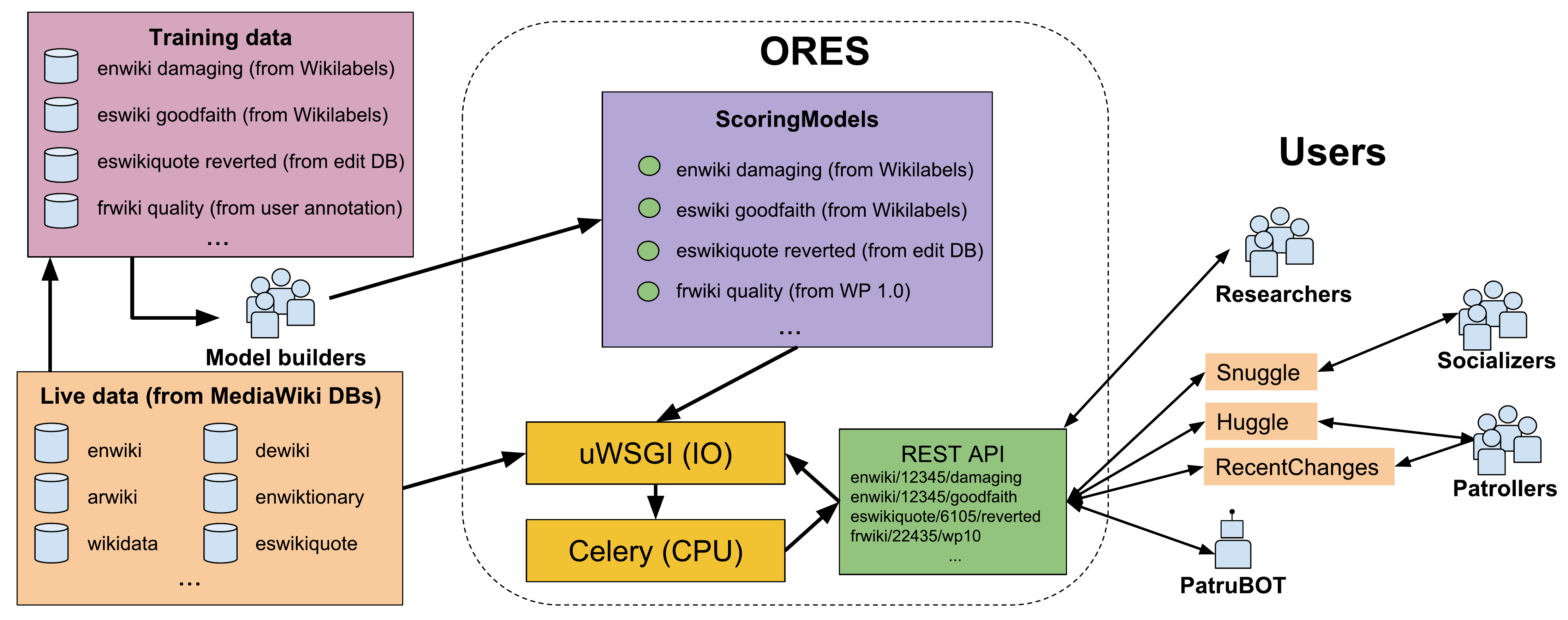}
  \caption{ORES conceptual overview.  Model builders design process for training ScoringModels from training data.  ORES hosts ScoringModels and makes them available to researchers and tool developers.}
  \label{fig:ores_data_user}
\end{figure*}

From a user's point of view (who is often a developer or a researcher), ORES is a collection of models that can be applied to Wikipedia content on demand. A user submits a query to ORES like ``is the edit identified by 123456 in English Wikipedia damaging?" (\url{https://ores.wikimedia.org/v3/scores/enwiki/123456/damaging}). ORES responds with a JSON \emph{score document} that includes a prediction (\texttt{false}) and a confidence level (\texttt{0.90}), which suggests the edit is probably not damaging.  Similarly, the user could ask for the quality level of the version of the article created by the same edit (\url{https://ores.wikimedia.org/v3/scores/enwiki/123456/articlequality}). ORES responds with a emph{score document} that includes a prediction (\texttt{Stub} -- the lowest quality class) and a confidence level (\texttt{0.91}), which suggests that the page needs a lot of work to attain high quality.  It is at this API interface that a user can decide what to do with the prediction: put it in a user interface, build a bot around it, use it as part of an analysis, or audit the performance of the models.

In the rest of this section, we discuss the strictly technical components of the ORES service.  See also Appendix~\ref{sec:appendix} for details about how we manage open source, self-documenting model pipelines, how we maintain a version history of the models deployed in production, and measurements of ORES usage patterns.  In the section~\ref{sec:ores_the_socio-technical_system}, we discuss socio-technical affordances we designed in collaboration with ORES' users.  
}

\subsection{Score documents}
\label{sec:score_documents}
The predictions made by ORES are human- and machine-readable.\footnote{Complex JSON documents are not typically considered ``human-readable,'' but score documents have a relatively short and consistent format, such that we have encountered several users with little to no programming experience querying the API manually and reading the score documents in their browser.}  In general, our classifiers will report a specific prediction along with a set of probability (likelihood) for each class.  By providing detailed information about a prediction, we allow users to re-purpose the prediction for their own use.  Consider the article quality prediction output in Figure~\ref{fig:english_wp10_prediction}.

\begin{figure}[h!]
        \makebox{\hrulefill}{
        \small
        \begin{verbatim}
  "score": {
    "prediction": "Start",
    "probability": {
      "FA": 0.00, "GA": 0.01, "B": 0.06, "C": 0.02, "Start": 0.75, "Stub": 0.16 }  }
        \end{verbatim}
        \hrule
        \normalsize}
        \caption{A score document -- the result of \url{https://ores.wikimedia.org/v3/scores/enwiki/34234210/articlequality}}
        \label{fig:english_wp10_prediction}
\end{figure}

A developer making use of a prediction like this in a user-facing tool may choose to present the raw prediction ``Start'' (one of the lower quality classes) to users or to implement some visualization of the probability distribution across predicted classes (75\% Start, 16\% Stub, etc.).  They might even choose to build an aggregate metric that weights the quality classes by their prediction weight (e.g. Ross's student support interface~\cite{ross2016visualizing} discussed in section~\ref{sec:ross_recommendations} or the \emph{weighted sum} metric from~\cite{halfaker2017interpolating}).

\subsection{Model information}
\label{sec:model_information}
In order to use a model effectively in practice, a user needs to know what to expect from model performance. For example, a \emph{precision} metric helps users think about how often is it that when an edit is predicted to be ``damaging,'' it actually is. Alternatively, a \emph{recall} metric helps users think about what proportion of damaging edits they should expect will be caught by the model. The target metric of an operational concern depends strongly on the intended use of the model.  Given that our goal with ORES is to allow people to experiment with the use and to appropriate prediction models in novel ways, we sought to build a general model information strategy.

\begin{figure}[htbp]
        \makebox{\hrulefill}{
        \small
        \begin{verbatim}
"damaging": {
  "type": "GradientBoosting",  "version": "0.4.0",
  "environment": {"machine": "x86_64", ...},
  "params": {"labels": [true, false], "learning_rate": 0.01, "min_samples_leaf": 1...},
  "statistics": {
    "counts": {
      "labels": {"false": 18702, "true": 743}, "n": 19445,
      "predictions": {
        "false": {"false": 17989, "true": 713},
        "true": {"false": 331, "true": 412}}},
    "precision": {"macro": 0.662, "micro": 0.962,
      "labels": {"false": 0.984, "true": 0.34}},
    "recall": {"macro": 0.758, "micro": 0.948,
      "labels": {"false": 0.962, "true": 0.555}},
    "pr_auc": {"macro": 0.721, "micro": 0.978,
      "labels": {"false": 0.997, "true": 0.445}},
    "roc_auc": {"macro": 0.923, "micro": 0.923,
      "labels": {"false": 0.923, "true": 0.923}},  ...  }}
        \end{verbatim}
        \hrule
        \normalsize}
        \caption{Model information for an English Wikipedia damage detection model -- the result of \url{https://ores.wikimedia.org/v3/scores/enwiki/?model_info&models=damaging}}
        \label{fig:english_damaging_model_info}
\end{figure}

The output captured in Figure~\ref{fig:english_damaging_model_info} shows a heavily trimmed (with ellipses) JSON output of \emph{model\_info} for the ``damaging'' model in English Wikipedia. What remains gives a taste of what information is available. There is structured data about what kind of algorithm is being used to construct the estimator, how it is parameterized, the computing environment used for training, the size of the train/test set, the basic set of fitness metrics, and a version number for secondary caches. A developer or researcher using an ORES model in their tools or analyses can use these fitness metrics to make decisions about whether or not a model is appropriate and to report to users what fitness they might expect at a given confidence threshold.

{\color{\highlightcolor}
\subsection{Scaling and robustness}
To be useful for Wikipedians and tool developers, ORES uses distributed computation strategies to provide a robust, fast, high-availability service.  Reliability is a critical concern, as many tasks are time sensitive. Interruptions in Wikipedia's algorithmic systems for patrolling have historically led to increased burdens for human workers and a higher likelihood that readers will see vandalism~\cite{geiger2013levee}. ORES also needs to scale to be able to be used in multiple different tools across different language Wikipedias. 

This horizontal scalability\footnote{C.f. \url{https://en.wikipedia.org/wiki/Scalability\#Horizontal_(Scale_Out)_and_Vertical_Scaling_(Scale_Up)}} is achieved in two ways: input-output (IO) workers (uwsgi\footnote{\url{https://uwsgi-docs.readthedocs.io/}}) and the computation (CPU) workers (celery\footnote{\url{http://www.celeryproject.org/}}).  Requests are split across available IO workers, and all necessary data is gathered using external APIs (e.g. the MediaWiki API\footnote{\url{http://enwp.org/:mw:MW:API}}).  The data is then split into a job queue managed by \emph{celery} for the CPU-intensive work.  This efficiently uses available resources and can dynamically scale, adding and removing new IO and CPU workers in multiple datacenters as needed.  This is also fault-tolerant, as servers can fail without taking down the service as a whole.

\subsubsection{Real-time processing}
The most common use case of ORES is real-time processing of edits to Wikipedia immediately after they are made. Those using patrolling tools like Huggle to monitor edits in real-time need scores available in seconds of when the edit is saved. We implement several strategies to optimize different aspects of this request pattern.

\leadin{Single score speed}
In the worst case scenario, ORES generates a score from scratch, as for scores requested by real-time patrolling tools. We work to ensure the median score duration is around 1 second, so counter-vandalism efforts are not substantially delayed (c.f. \cite{geiger2013levee}).  Our metrics show for the week April 6-13th, 2018, our median, 75\%, and 95\% percentile response timings are 1.1, 1.2, and 1.9 seconds respectively.  This includes the time to process the request, gather data, process data into features, apply the model, and return a score document.  We achieve this through computational optimizations to ensure data processing is fast, and by choosing not to adopt new features and modeling strategies that might make score responses too slow (e.g., the RNN-LSTM strategy used in~\cite{dang2017end} is more accurate than our article quality models, but takes far longer to compute).  

\leadin{Caching and precaching}
In order to take advantage of our users' overlapping interests in scoring recent activity, we maintain a least-recently-used (LRU) cache\footnote{Implemented natively by Redis, \url{https://redis.io}} using a deterministic score naming scheme (e.g. \texttt{enwiki:123456:damaging} would represent a score needed for the English Wikipedia damaging model for the edit identified by 123456).  This allows requests for scores that have recently been generated to be returned within about 50ms -- a 20X speedup. To ensure scores for \emph{all recent edits} are available in the cache for real-time use cases, we implement a ``precaching'' strategy that listens to a high-speed stream of recent activity in Wikipedia and automatically requests scores for the subset of actions that are relevant to a model.  With our LRU and precaching strategy, we consistently attain a cache hit rate of about 80\%.

\leadin{De-duplication}
In real-time ORES use cases, it is common to receive many requests to score the same edit/article right after it was saved.  We use the same deterministic score naming scheme from the cache to identify scoring tasks, and ensure that simultaneous requests for that same score are de-duplicated.  This allows our service to trivially scale to support many different robots and tools requesting scores simultaneously on the same wiki.

\subsubsection{Batch processing}
In our logs, we have observed bots submitting large batch processing jobs to ORES once per day. Many different types of Wikipedia's bots rely on periodic, batch processing strategies to support Wikipedian work~\cite{geiger2011lives}. Many bots build daily or weekly worklists for Wikipedia editors (e.g. \cite{cosley2007suggestbot}). Many of these tools have adopted ORES to include an article quality prediction for use in prioritization\footnote{See complete list: \url{http://enwp.org/:mw:ORES/Applications}}). Typically, work lists are either built from all articles in a Wikipedia language version (>5m in English) or from some subset of articles specific to a single WikiProject (e.g. WikiProject Women Scientists claims about 6k articles\footnote{As demonstrated by \url{https://quarry.wmflabs.org/query/14033}}). Also, many researchers are using ORES for analyses, which presents as a similar large burst of requests.  

In order to most efficiently support this type of querying activity, we implemented batch optimizations by splitting IO and CPU operations into distinct stages.  During the IO stage, all data is gathered for all relevant scoring jobs in batch queries.  During the CPU stage, scoring jobs are split across our distributed processing system discussed above.  This batch processing enables up to a 5X increase in speed of response for large requests\cite{sarabadani2017building}.  At this rate, a user can request tens of millions of scores in less than 24 hours in the worst case scenario (no scores were cached), without substantially affecting the service for others.
}

\section{ORES the sociotechnical system}
\label{sec:ores_the_sociotechnical_system}
\label{sec:ores_the_socio-technical_system}
{\color{\highlightcolor}
While ORES can be considered as a purely technical information system (e.g. as bounded in Fig.~\ref{fig:ores_data_user}) and discussed in the prior section, it is useful to take a broader view of ORES as a socio-technical system that is maintained and operated by people in certain ways and not others. Such a move is in line with recent literature encouraging a shift from focusing on ``algorithms" specifically to ``algorithmic systems"~\cite{seaver2017algorithms}.  In this broader view, the role of our engineering team is particularly relevant, as ORES is not a fully-automated, fully-decoupled ``make your own model" system, as in the kind of workflow exemplified by Google's Teachable Machine.\footnote{\url{https://www.blog.google/technology/ai/teachable-machine/}} In the design of ORES, this could have been a possible configuration, in which Wikipedians would have been able to upload their own training datasets, tweak and tune various parameters themselves, then hit a button that would deploy the model on the Wikimedia Foundation's servers to be served publicly at scale via the API --- all without needing approval or actions by us. 

Instead, we have played a somewhat stronger gatekeeping role than we initially envisioned, as no new models can be deployed without our express approval. However, our team has long operated with a particular orientation and set of values, most principally committing to collaboratively build ML models that contributors at local Wikimedia projects believe may be useful in supporting their work. Our team --- which includes paid staff at the Wikimedia Foundation, as well as some part-time volunteers --- has long acted more like a service or consulting unit than a traditional product team. We provide customized resources after extended dialogue with representatives from various wiki communities who have expressed interest in using ML for various purposes.

\subsection{Collaborative model design}
\label{sec:collaborative_model_design}
From the perspective of Wikipedians who want to add a new model/classifier to ORES, the first step is to request a model. They message the team and outline what kind of model/classifier they would like and/or what kinds of problems they would like to help tackle with ML. Requesting a classifier and discussing the request take place on the same communication channels used by active contributors of Wikimedia projects. The team has also performed outreach, including giving presentations to community members virtually and at meetups\footnote{For example, see \url{https://w.wiki/SRc} ``Using AI to Keep Wikipedia Open" and \url{https://w.wiki/SRd} ``The People's Classifier: Towards an open model for algorithmic infrastructure"}. Wikimedia contributors have also spread news about ORES and our team's services to others through word-of-mouth.  For example, the Portuguese Wikipedia demonstrative case we describe in section~\ref{sec:portuguese_wikipedia_article_quality} was started because of a presentation that someone outside of our team gave about ORES at WikiCon Portugual.\footnote{\url{https://meta.wikimedia.org/wiki/WikiCon_Portugal}}

Even though our team has a more collaborative, participatory, and consulting-style approach, we still play a strong role in the design, scoping, training, and deployment of ORES' models. Our team assesses the feasibility of each request, works with requester(s) to define and scope the proposed model, helps collect and curate labeled training data, and helps engineer features. We often go through multiple rounds of iteration based on the end users' experience with the model and we help community members think about how they might integrate ORES scores within existing or new tools to support their work. ORES could become a rather different kind of socio-technical system if our team performed this exact same kind of work in the exact same technical system, but had different values, approaches, and procedures.

\subsubsection{Trace extraction and manual labeling}
Labeled data is the heart of any model, as ML can only be as good as the data used to train it \cite{geiger2020garbage}. It is through focusing on labeled observations that we help requesters understand and negotiate the \emph{meaning to be modeled} \cite{jacobs2020meaning} by an ORES classifier. We employ two strategies around labeling: found trace data and community labeling campaigns.

\leadin{Found trace data from wikis}
Wikipedians have long used the wiki platform to record digital traces about decisions they make~\cite{geiger2011trace}. These traces can sometimes be assumed to reflect a useful labeled data for modeling, although like all found data, this must be done with much care. One of the most commonly requested models are classifiers for the quality of an article at a point in time. The quality of articles is an important concept across language versions, which are independently written according to their own norms, policies, and procedures, and standards. Yet most Wikipedia language versions have a more-or-less standardized process for reviewing articles. Many of these processes began as ways to select high quality articles to feature on their wiki's home page as the ``article of the day.'' Articles that do not pass are given feedback; like in academic peer review, there can be many rounds of review and revision. In many wikis, an article can be given a range of scores, with specific criteria defining each level.

Many language versions have developed quality scales that formalize their concepts of article quality~\cite{stvilia2009issues}. Each local language version can have their own quality scale, with their own standards and assessment processes (e.g. English Wikipedia has a 7-level scale, Italian Wikipedia has a 5-level scale, Tamil Wikipedia has a 3-level scale). Even wikis that use the same scales can have different standards for what each level means and that meaning can change over time~\cite{halfaker2017interpolating}. 

Most wikis also have standard ``templates" for leaving assessment trace data. In English Wikipedia and many others, these templates are placed by WikiProjects (subject-focused working groups) on the ``Talk pages" that are primarily used for discussing the article's content. For wikis that have these standardized processes, scales, and trace data templates, our team asks the requesters of article quality models to provide some information and links about the assessment process. The team uses this to build scripts that scrape this into training datasets.  This process is highly iterative, as processing mistakes and misunderstandings about the meaning and historical use of a template often need to be worked out in consultations with Wikipedians who are more well versed in their own community's history and processes. These are reasons why it is crucial to involve community members in ML throughout the process.

\begin{figure}[h]
  \centering
  \includegraphics[width=.50\textwidth]{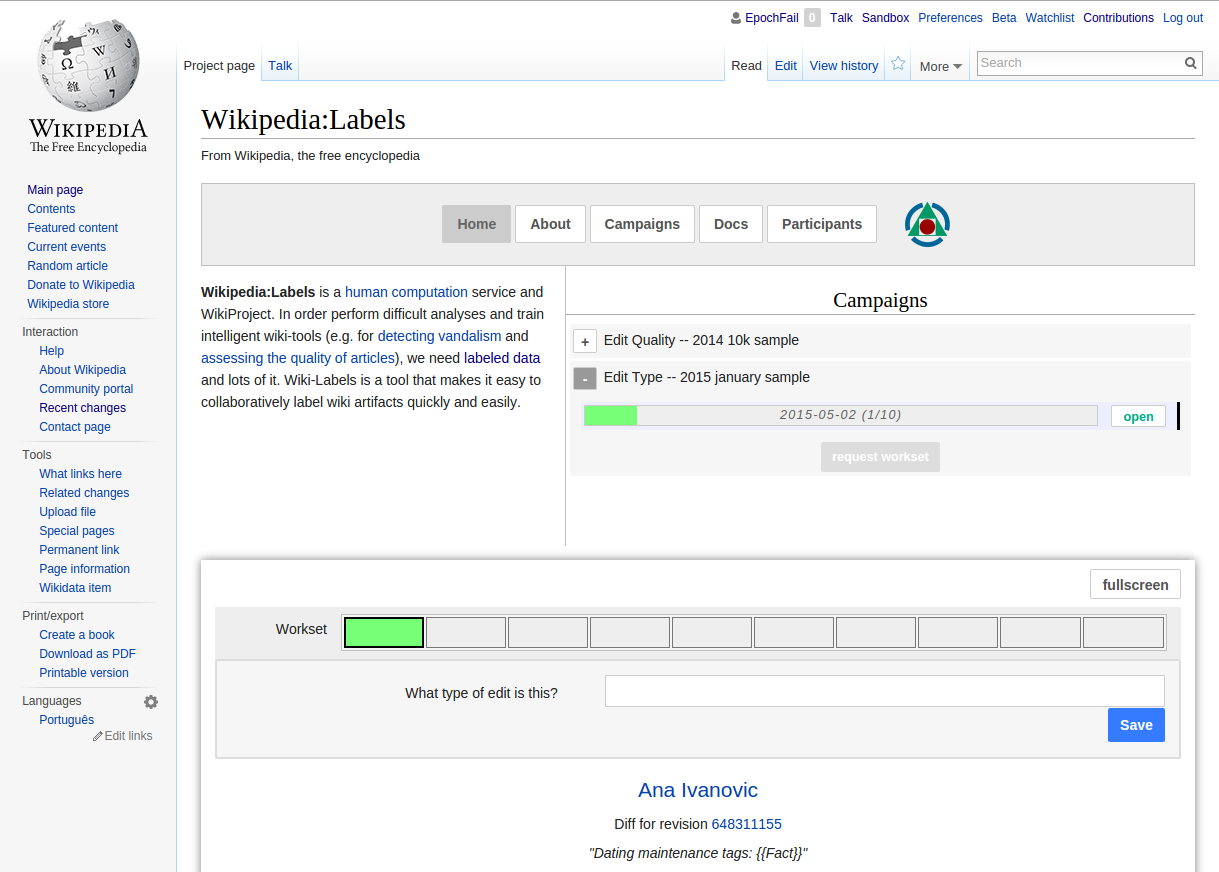}
  \caption{The Wiki labels interface embedded in Wikipedia}
  \label{fig:wikilabels_screenshot}
\end{figure}

\leadin{Manual labeling campaigns with Wiki Labels}
In many cases, we do not have any suitable trace data we can extract as labels. For these cases, we ask our Wikipedian collaborators to perform a community labeling exercise. This can have high cost: for some models, we need tens of thousands of observations in order to achieve high fitness and adequately test performance.  To minimize that cost, we developed a high-speed, collaborative labeling interface called ``Wiki Labels.''\footnote{\url{http://enwp.org/:m:Wiki labels}}  We work with model requesters to design an appropriate sampling strategy for items to be labeled and appropriate labeling interfaces, load the sample of items into the system, and help requesters recruit collaborators from their local wiki to generate labels. Labelers to request ``worksets" of observations to label, which are presented in a customizable interface. Different campaigns require different ``views" of the observation to be labeled (e.g., a whole page, a specific sentence within a page, a sequence of edits by an editor, a single edit, a discussion post, etc.) and ``forms" to capture the label data (e.g., ``Is this edit damaging?", ``Does this sentence need a citation?", ``What quality level is this article?", etc.). Unlike with most Wikipedian edit review interfaces, Wiki Labels does not show information about the user who made the edit, to help mitigate implicit bias against certain kinds of editors (although this does not mitigate biases against certain kinds of content).

Labeling campaigns may be labor-intensive, but we have found they often prompt us to reflect and specify what precisely it is the requesters want to predict. Many issues can arise in ML from overly broad or poorly operationalized theoretical constructs \cite{jacobs2020meaning}, which can play out in human labeling \cite{geiger2020garbage}. For example, when we first started discussing modeling patrolling work with Wikipedians, it became clear that these patrollers wanted to give vastly different responses when they encountered different kinds of ``damaging'' edits.  Some of these ``damaging'' edits were clearly seen as ``vandalism'' to patrollers, where the editor in question appears to only be interested in causing harm.  In other cases, patrollers encountered ``damaging'' edits that they felt certainly lowered the quality of the article, but were made in a way that they felt was more indicative of a mistake or misunderstanding. 

Patrollers felt these second kind of damaging edits were from people who were generally trying to contribute productively, but they were still violating some rule or norm of Wikipedia. Wikipedians have long referred to this second type of ``damage'' as ``good-faith,'' which is common among new editors and requires a carefully constructive response. ``Good-faith'' is a well-established term in Wikipedian culture\footnote{\url{https://enwp.org/WP:AGF}}, with specific local meanings that are different than their broader colloquial use --- similar to how Wikipedians define ``consensus'' or ``neutrality''\footnote{See \url{https://enwp.org/WP:CONSENSUS} \& \url{https://enwp.org/WP:NPOV}}.  We used this understanding and the distinction it provided to build a \emph{form} for Wiki Labels that allowed Wikipedians to distinguish between these cases.  That allowed us to build two separate models which allow users to filter for edits that are likely to be good-faith mistakes ~\cite{halfaker2017automated}, to just focus on vandalism, or to apply themselves broadly to all damaging edits.

\subsubsection{Demonstrative case: Article quality in Portuguese vs Basque Wikipedia}
\label{sec:portuguese_wikipedia_article_quality}
In this section, we discuss how we worked with community members from Portuguese Wikipedia to develop article quality models based on trace data, then we contrast with what we did for Basque Wikipedia where no trace data was available. These demonstrative cases illustrate a broader pattern we follow when collaboratively building models.  We obtained consent from all editors mentioned to share their story with their usernames, and they have reviewed this section to verify our understanding of how and why they worked with us. 

GoEThe, a volunteer editor from the Portuguese Wikipedia, attended a talk at WikiCon Portugal where a Wikimedia Foundation staff member (who was not on our team) mentioned ORES and the potential of article quality models to support work. GoEThe then found our documentation page for requesting new models \footnote{\url{https://www.mediawiki.org/wiki/ORES/Get_support}}.  He clicked on a blue button titled ``Request article quality model''.  This led to a template for creating a task in our ticketing system.\footnote{\url{https://phabricator.wikimedia.org}}  The article quality template asks questions like ``How do Wikipedians label articles by their quality level?" and ``What levels are there and what processes do they follow when labeling articles for quality?"  These are good starting points for a conversation about what article quality is and how to measure it at scale.

Our engineers responded by asking follow-up questions about the digital traces Portuguese Wikipedians have long used to label articles by quality, which include the history of the bots, templates, automation, and peer review used to manage the labeling process. As previously discussed, existing trace data from decisions made by humans can appear to be a quick and easy way to get labels for training data, but it can also be problematic to rely on these traces without understanding the conditions of their production. After an investigation, we felt more confident there was consistent application of a well-defined quality scale\footnote{\url{https://pt.wikipedia.org/wiki/Predefini\%C3\%A7\%C3\%A3o:Escala_de_avalia\%C3\%A7\%C3\%A3o}} and labels were consistently applied using a common template named ``Marca de projeto".  Meanwhile, we encountered another volunteer (Chtnnh) who was not a Portuguese Wikipedian, but interested in ORES and machine learning. Chtnnh worked with our team to iterate on scripts for extracting the traces with increasing precision.  At the same time, a contributor to Portuguese Wikipedia named He7d3r joined, adding suggestions about our data extraction methods and contributing code.  One of He7d3r's major contributions were features based on a Portuguese ``words\_to\_watch" list --- a collection of imprecise or exaggerated words, like vision\'ario (visionary) and brilhante (brilliant).\footnote{\url{https://pt.wikipedia.org/wiki/Wikip\%C3\%A9dia:Palavras\_a\_evitar}}. We even gave He7d3r access to one of our servers, so that he could more easily experiment with extracting labels, engineering new features, and building new models.

In Portuguese Wikipedia, there was already trace data available that was suitable for training and testing an article quality model, which is not always the case.  For example, we were approached about an article quality model by another volunteer from Basque Wikipedia -- a much smaller language community -- which had no history of labeling articles and thus no trace data we could use as observations.  However, they had drafted a quality scale and were hoping that we could help them get started by building a model for them.  In this case, we worked with a volunteer from Basque Wikipedia to develop heuristics for quality (e.g. article length) and used those heuristics to build a stratified sample, which we loaded into the \emph{Wiki Labels} tool with a \emph{form} that matched their quality scale.  Our Basque collaborator also gave direct recommendations on our feature engineering.  For example, after a year of use, they noticed that the model seemed to put too much weight on the presence of \emph{Category links}\footnote{\url{https://en.wikipedia.org/wiki/Help:Category}} in the articles. They requested we remove the features and retrain the model.  While this produced a small, measurable drop in our formal fitness statistics, editors reported that the new model matched their understanding of quality better in practice.

These two cases demonstrate the process by which volunteers --- many of whom had no to minimal experiences with software development and ML --- work with us to encode their understanding of quality and the needs of their work practices into a modeling pipeline. These cases also show how there is far more socio-technical work involved in the design, training, and optimization of machine learning models than just labeling training data.  Volunteers play a critical role in deciding that a model is necessary, in defining the output classes the model should predict, in associating labels with observations (either manually or by interpreting historic trace data), and in engineering model features.  In some cases, volunteers rely on our team to do most of the software and model engineering work, then they mainly give us feedback about model performance, but more commonly, the roles that volunteers take on overlap significantly with us researchers and engineers.  The result is a process that is collaborative and is similar in many ways to the open collaboration practice of Wikipedia editors.  
}

\subsubsection{Demonstrative case: Italian Wikipedia thematic analysis}
\label{sec:italian_wikipedia_thematic_analysis}
Italian Wikipedia was one of the first wikis outside of English Wikipedia where we deployed models for helping editors detect vandalism.  After we deployed the initial version of the model, we asked Rotpunkt --- our local collaborator who originally requested we build the model and helped us develop language-specific features --- to help us gather feedback about how the model was performing in practice.  He put together a page on Italian Wikipedia\footnote{\url{https://it.wikipedia.org/wiki/Progetto:Patrolling/ORES}} and encouraged patrollers to note the mistakes that the model was making there.  He created a section for reporting ``falsi positivi'' (false-positives).  Within several hours, Rotpunkt and others noticed trends in edits that ORES was getting wrong.  They sorted false positives under different headers, representing themes they were seeing --- effectively performing an audit of ORES through an inductive, grounded theory-esque thematic coding process.

One of the themes they identified was ``correzioni verbo avere'' (``corrections to the verb for \emph{have}'').  The word ``ha'' in Italian translates to the English verb ``to have''.  In English and many other languages, ``ha'' signifies laughing, which is not usually found in encyclopedic prose. Most non-English Wikipedias receive at least some  English vandalism like this, so we had built a common feature in all patrolling support models called ``informal words'' to capture these patterns. Yet in this case, ``ha'' should not carry signal of damaging edits in Italian, while ``hahaha'' still should. Because of Rotpunkt and his collaborators in Italian Wikipedia, we were recognized the source of this issue, to removed ``ha'' from that informal list for Italian Wikipedia, and deployed a model that showed clear improvements.

This case demonstrates the innovative way our Wikipedian collaborators have advanced their own processes for working with us.  It was their idea to group false positives by theme and characteristics, which made for powerful communication.  Each theme identified by the Wikipedians was a potential bug somewhere in our model. We may have never found such a bug without the specific feedback and observations that end-users of the model were able to give us.

{\color{\highlightcolor}
\subsection{Technical affordances for adoption and appropriation}
Our engineering team values being responsive to community needs. To this end, we designed ORES as a technical system in a way that we can more easily re-architect or extend the system in response to initially unanticipated needs.  In some cases, these needs only became apparent after the base system (described in section~\ref{sec:ores_the_technology}) was in production --- where communities we supported were able to actively explore what might be possible with ML in their projects, then raised questions or asked for features that we had not considered. Through maintaining close relationships with the communities we support, we have been able to extend and adapt the technical systems in novel ways to support their use of ORES. We identified and implemented two novel affordances that support the adoption and re-appropriation of ORES models: dependency injection and threshold optimization.  

\subsubsection{Dependency injection}
When we originally developed ORES, we designed our feature engineering strategy based on a dependency injection framework\footnote{\url{https://en.wikipedia.org/wiki/Dependency_injection}}.  A specific feature used in prediction (e.g., \emph{number of references}) depends on one or more datasources (e.g. \emph{article text}).  Many different features can depend on the same datasource.  A model uses a sampling of features in order to make predictions.  A dependency solver allowed us to efficiently and flexibly gather and process the data necessary for generating the features for a model --- initially a purely technical decision.

After working with ORES' users, we received requests for ORES to generate scores for edits before they were saved, as well as to help explore the reasons behind some of the predictions.  After a long consultation, we realized we could provide our users with direct access to the features that ORES used for making predictions and let those users \emph{inject} features and even the datasources they depend on.  A user can gather a score for an edit or article in Wikipedia, then request a new scoring job with one of those features or underlying datasources modified to see how the prediction would change. For example, how does ORES differently judge edits from unregistered (anon) vs registered editors? Figure~\ref{fig:anon_injection} demonstrates two prediction requests to ORES with features injected. 

\begin{figure*}[h]
\centering
\begin{subfigure}[t]{.5\textwidth}
  \makebox{\hrulefill}{
  \small
  \begin{verbatim}
  "damaging": {
    "score": {
      "prediction": false,
      "probability": {
        "false": 0.938910157824447,
        "true": 0.06108984217555305   }   }  }
  \end{verbatim}
  \hrule
  \normalsize}
  \caption{Prediction with \texttt{anon = false} injected}
  \label{fig:anon_injection_false}
\end{subfigure}~~
\begin{subfigure}[t]{.5\textwidth}
  \makebox{\hrulefill}{
  \small
  \begin{verbatim}
  "damaging": {
    "score": {
      "prediction": false,
      "probability": {
        "false": 0.9124151990561908,
        "true": 0.0875848009438092   }   }   }
  \end{verbatim}
  \hrule
  \normalsize}
  \caption{Prediction with \texttt{anon = true} injected}
  \label{fig:anon_injection_true}
\end{subfigure}
\caption{Two ``damaging'' predictions about the same edit are listed for ORES.  In one case, ORES scores the prediction assuming the editor is unregistered (anon) and in the other, ORES assumes the editor is registered.}
\label{fig:anon_injection}
\end{figure*}

Figure~\ref{fig:anon_injection_false} shows that ORES' ``damaging'' model concludes the edit is not damaging with 93.9\% confidence. Figure~\ref{fig:anon_injection_true} shows the prediction if the edit were saved by an anonymous editor.  ORES would still conclude that the edit was not damaging, but with less confidence (91.2\%).  By following a pattern like this, we better understand how ORES prediction models account for anonymity with practical examples.  End users of ORES can inject raw text of an edit to see the features extracted and the prediction, without making an edit at all. 
}

\subsubsection{Threshold optimization}
\label{sec:appendix.threshold_optimization}
When we first started developing ORES, we realized that operational concerns of Wikipedia's curators need to be translated into confidence thresholds for the prediction models.  For example, counter-vandalism patrollers seek to catch all (or almost all) vandalism quickly.  That means they have an operational concern around the \emph{recall} of a damage prediction model.  They would also like to review as few edits as possible in order to catch that vandalism.  So they have an operational concern around the \emph{filter-rate}---the proportion of edits that are not flagged for review by the model~\cite{halfaker2016notes}. By finding the threshold of prediction confidence that optimizes the filter-rate at a high level of recall, we can provide patrollers with an effective trade-off for supporting their work.  We refer to these optimizations as \emph{threshold optimizations} and ORES provides information about these thresholds in a machine-readable format so tool developers can write code that automatically detects the relevant thresholds for their wiki/model context.

Originally when we developed ORES, we defined these threshold optimizations in our deployment configuration, which meant we would need to re-deploy the service any time a new threshold was needed. We soon learned users wanted to be able to search through fitness metrics to choose thresholds that matched their own operational concerns on demand.  Adding new optimizations and redeploying became a burden on us and a delay for our users, so we developed a syntax for requesting an optimization from ORES in real-time using fitness statistics from the model's test data. For example, \texttt{maximum recall @ precision >= 0.9} gets a useful threshold for a patrolling auto-revert bot or \texttt{maximum filter\_rate @ recall >= 0.75} gets a useful threshold for patrollers who are filtering edits for human review.

\begin{figure}[htbp]
        \makebox{\hrulefill}{
        \small
        \begin{verbatim}
  {"threshold": 0.32, ...,
   "filter_rate": 0.89, "fpr": 0.087,  "precision": 0.23, "recall": 0.75}
        \end{verbatim}
        \hrule
        \normalsize}
        \caption{A threshold optimization -- the result of \url{https://ores.wikimedia.org/v3/scores/enwiki/?models=damaging&model_info=statistics.thresholds.true.'maximum filter_rate @ recall >= 0.75'}}
        \label{fig:english_damaging_threshold_optimization}
\end{figure}

Figure ~\ref{fig:english_damaging_threshold_optimization} shows that, when a threshold is set on 0.299 likelihood of \texttt{damaging=true}, a user can expect to get a recall of 0.751, precision of 0.215, and a filter-rate of 0.88.  While the precision is low, this threshold reduces the overall workload of patrollers by 88\% while still catching 75\% of (the most egregious) damaging edits.

\subsubsection{Demonstrative case: Ross' recommendations for student authors}
\label{sec:ross_recommendations}

One of the most noteworthy (and initially unanticipated) applications of ORES to support new editors is the suite of tools developed by Sage Ross to support the Wiki Education Foundation's\footnote{\url{https://wikiedu.org/}} activities.  Their organization supports classroom activities that involve editing Wikipedia.  They develop tools and dashboards that help students contribute successfully and to help teachers monitor their students' work.  Ross published about how he interprets meaning from ORES article quality models~\cite{ross2016visualizing} (an example of re-appropriation) and how he has used the article quality model in their student support dashboard\footnote{\url{https://dashboard-testing.wikiedu.org}} in a novel way.  Ross's tool uses our dependency injection system to suggest work to new editors.  This system asks ORES to score a student's draft article, then asks ORES to reconsider the predicted quality level of the article with \emph{one more header}, \emph{one more image}, \emph{one more citation}, etc. --- tracking changes in the prediction and suggesting the largest positive change to the student.  In doing so, Ross built an intelligent user interface that can expose the internal structure of a model in order to recommend the most productive change to a student's article --- the change that will most likely bring it to a higher quality level. 

{\color{\highlightcolor}
This adoption pattern leverages ML to support an under-supported user class, as well as balances concerns around quality control efficiency and newcomer socialization with a completely novel strategy.  By helping student editors learn to structure articles to match Wikipedians' expectations, Ross's tool has the potential to both improve newcomer socialization and minimize the burden for Wikipedia's patrollers --- a dynamic that has historically been at odds~\cite{halfaker2013rise, teblunthuis2018revisiting}.  By making it easy to surface how a model's predictions vary based on changes to input, Ross was able to provide a novel functionality we had never considered.

\subsubsection{Demonstrative case: Optimizing thresholds for RecentChanges Filters}
\label{sec:recent_changes_filters}
In October of 2016, the Global Collaboration Team at the Wikimedia Foundation began work to redesign Wikipedia's RecentChanges feed,\footnote{\url{https://www.mediawiki.org/wiki/Edit_Review_Improvements/New_filters_for_edit_review}} a tool used by Wikipedia editors to track edits to articles in near real-time, which is used for patrolling edits for damage, to welcome new editors, to review new articles for their suitability, and to otherwise stay up to date with what is happening in Wikipedia.  The most prominent feature of the redesign they proposed would bring in ORES's ``damaging'' and ``good-faith'' models as flexible filters that could be mixed and matched with other basic filters.  This would (among other use-cases) allow patrollers to more easily focus on the edits that ORES predicts as likely to be damaging.  However, a question quickly arose from the developers and designers of the tool: at what confidence level should edits be flagged for review?  

We consulted with the design team about the operational concerns of patrollers: that recall and filter-rate need to be balanced in order for effective patrolling~\cite{halfaker2016notes}.  The designers on the team identified the value of having a high precision threshold as well.  After working with them to define threshold and making multiple new deployment of the software, we realized that there was an opportunity to automate the threshold discovery process and allow the client (in this case, the MediaWiki software's RecentChanges functionality) to use an algorithm to select appropriate thresholds.  This was very advantageous for both us and the product team.  When we deploy an updated version of a model, we increment a version number.  MediaWiki checks for a change in that version number and re-queries for an updated threshold optimization as necessary.  

This case shows how we were able to formalize a relationship between model developers and model users.  Model users needed to be able to select appropriate confidence thresholds for the UX that they have targeted and they did not want to need to do new work every time that we make a change to a model. Similarly, we did not want to engage in a heavy consultation and iterative deployments in order to help end users select new thresholds for their purposes.  By encoding this process in ORES API and model testing practices, we give both ourselves and ORES users a powerful tool for minimizing the labor involved in appropriating ORES' models.
}

\subsubsection{Demonstrative case: Anonymous users and Tor users}
\label{sec:anonymous_and_tor_users}
\begin{figure*}[h!]
\centering
\begin{subfigure}[t]{.33\textwidth}
  \centering
  \includegraphics[width=.85\textwidth]{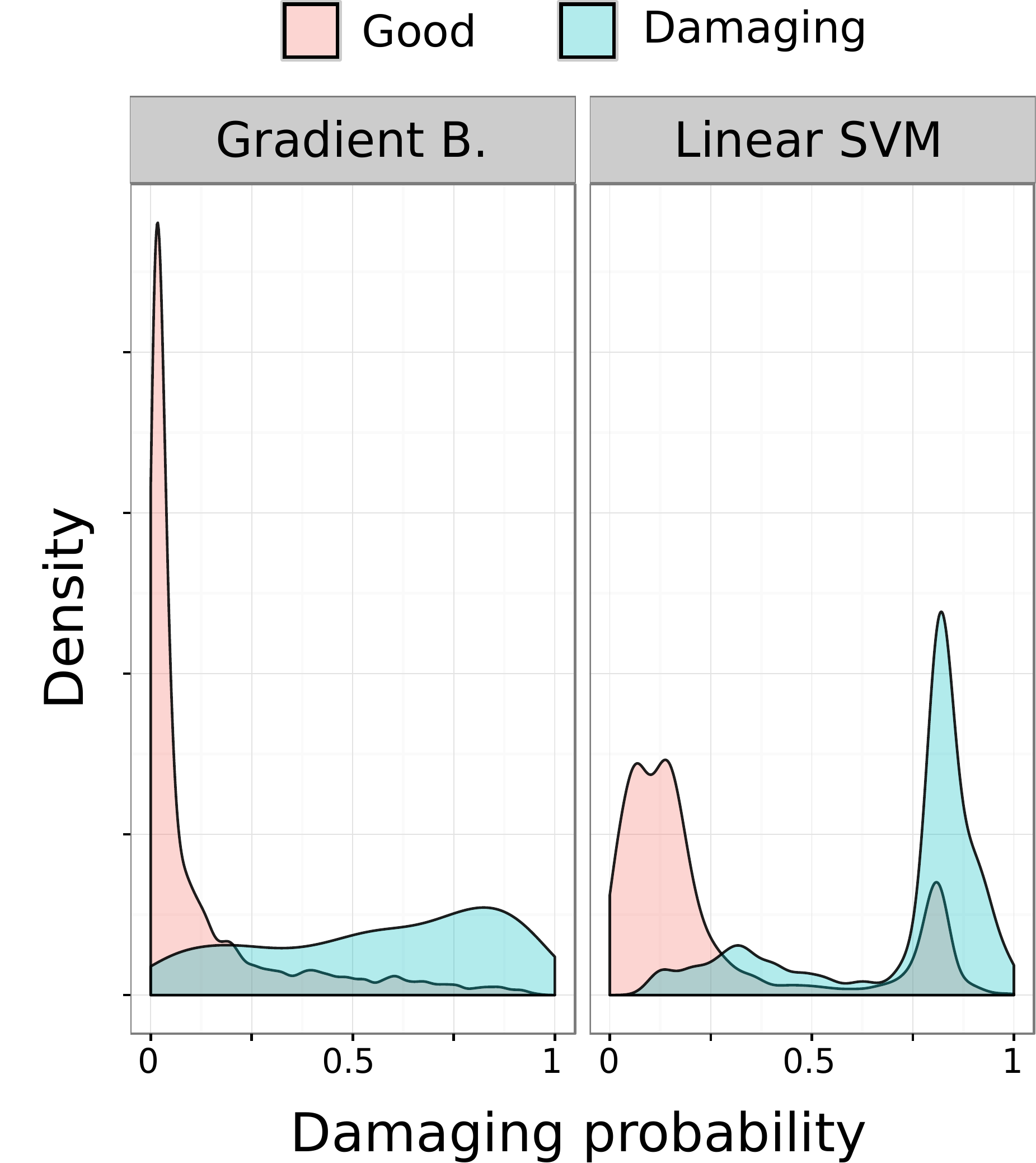}
  \caption{No injected features}
  \label{fig:natural_damaging_gb_bs_svc}
\end{subfigure}~~
\begin{subfigure}[t]{.33\textwidth}
  \centering
  \includegraphics[width=.85\textwidth]{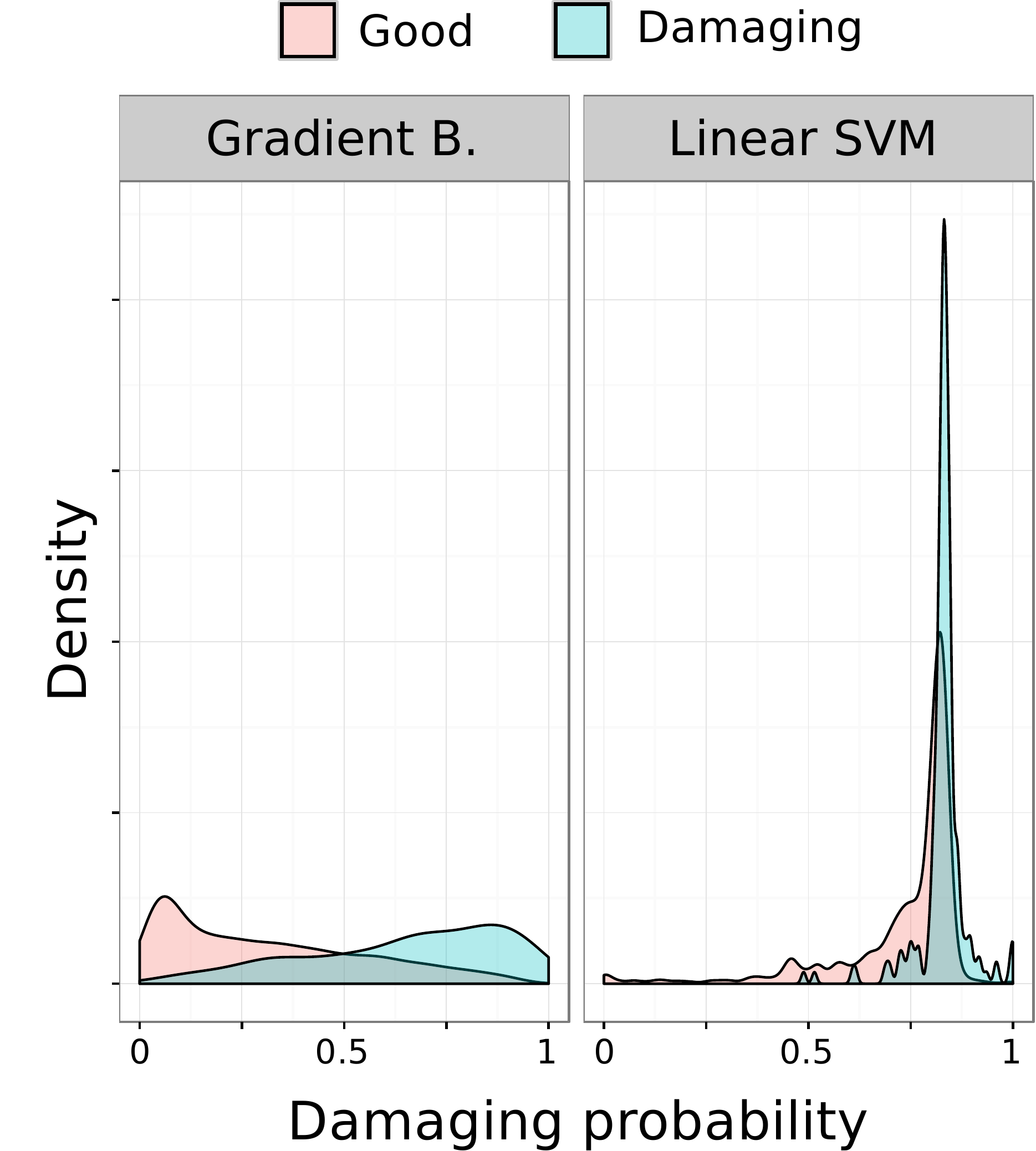}
  \caption{Everyone is anonymous}
  \label{fig:anon_damaging_gb_bs_svc}
\end{subfigure}~~
\begin{subfigure}[t]{.33\textwidth}
  \centering
  \includegraphics[width=.85\textwidth]{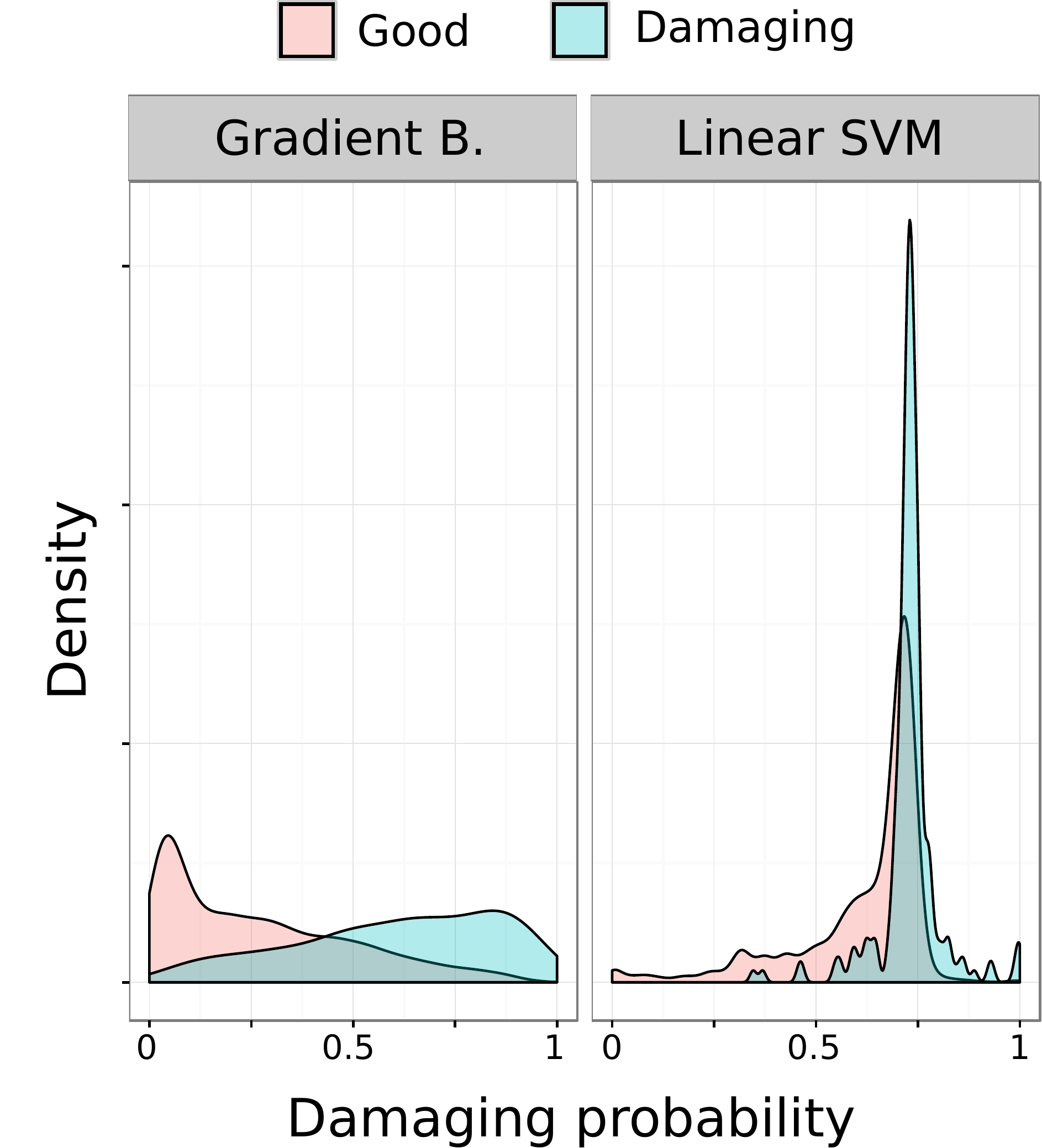}
  \caption{Everyone is newly registered}
  \label{fig:newcomer_damaging_gb_bs_svc}
\end{subfigure}
\caption{The distributions of the probability of a single edit being scored as ``damaging'' based on injected features for the target user-class is presented.  Note that when injecting user-class features (anon, newcomer), all other features are held constant.}
\label{fig:prediction_error_for_anons_and_newcomers}
\end{figure*}

Shortly after we deployed ORES, we received reports that ORES's damage detection models were overly biased against anonymous editors.  At the time, we were using Linear SVM\footnote{\url{http://scikit-learn.org/stable/modules/generated/sklearn.svm.LinearSVC.html}} estimators to build classifiers, and we were considering making the transition towards ensemble strategies like GradientBoosting and RandomForest estimators.\footnote{\url{http://scikit-learn.org/stable/modules/ensemble.html}}  We took the opportunity to look for bias in the error of estimation between anonymous editors and registered editors.  By using our dependency injection strategy, we could ask our current prediction models how they would change their predictions if the exact same edit were made by a different editor.

Figure~\ref{fig:prediction_error_for_anons_and_newcomers} shows the probability density of the likelihood of ``damaging'' given three different passes over the exact same test set, using two of our modeling strategies.  Figure~\ref{fig:natural_damaging_gb_bs_svc} shows that, when we leave the features to their natural values, it appears that both algorithms are able to learn models that differentiate effectively between damaging edits (high-damaging probability) and non-damaging edits (low-damaging probability) with the odd exception of a large amount of non-damaging edits with a relatively high-damaging probability around 0.8 in the case of the Linear SVM model.  Figures~\ref{fig:anon_damaging_gb_bs_svc} and \ref{fig:newcomer_damaging_gb_bs_svc} show a stark difference.  For the scores that go into these plots, characteristics of anonymous editors and newly registered editors were injected for all of the test edits.  We can see that the GradientBoosting model can still differentiate damage from non-damage while the Linear SVM model flags nearly all edits as damage in both cases.

Through the reporting of this issue and our analysis, we identified the weakness of our estimator and mitigated the problem.  Without a tight feedback loop, we likely would not have noticed how poorly ORES's damage detection models were performing in practice.  It might have caused vandal fighters to be increasingly (and inappropriately) skeptical of contributions by anonymous editors and newly registered editors---two groups that are already met with unnecessary hostility\footnote{\url{http://enwp.org/:en:Wikipedia:IPs_are_human_too}}\cite{halfaker2013rise}.

{\color{\highlightcolor}
This case demonstrates how the dependency injection strategy can be used in practice to empirically and analytically explore the bias of a model. Since we have started talking about this case publicly, other researchers have begun using this strategy to explore ORES predictions for other user groups.  Notably, Tran et al. used ORES' feature injection system to explore the likely quality of Tor users' edits\footnote{\url{https://en.wikipedia.org/wiki/Wikipedia:Advice_to_users_using_Tor}} as though they were edits saved by newly registered users~\cite{tran2019tor}.  We likely would not have thought to implement this functionality if it were not for our collaborative relationship with developers and researchers using ORES, who raised these issues.

\subsection{Governance and decoupling: delegating decision-making}
If ``code is law'' \cite{lessig1999code}, then internal decision-making processes of technical teams constitute governance systems. Decisions about what models should be built and how they should be used play major roles in how Wikipedia functions. ORES as a socio-technical system is variously \textit{coupled and decoupled}, which as we use it, is a distinction about how much vertical integration and top-down control there is over different kinds of labor involved in machine learning. Do the same people have control and/or responsibility over deciding what the model will classify, labeling training data, curating and cleaning labeled training data, engineering features, building models, evaluating or auditing models, deploying models, and developing interfaces and/or agents that use scores from models? A highly-coupled ML socio-technical system has the same people in control of this work, while a highly-decoupled system involves distributing responsibility for decision-making --- providing entrypoints for broader participation, auditing, and contestability \cite{mulligan2019shaping}. 

Our take on decoupling draws on feminist standpoint epistemology, which often critiques a single monolithic ``God's eye view'' or ``view from nowhere'' and emphasizes accounts from multiple positionalities \cite{haraway1988situated,harding1987feminism}. Similar calls drawing on this work have been made in texts like the Feminist Data Set Manifesto \cite[see][]{sinders2019making}, as more tightly coupled approaches enforce a particular singular top-down worldview. Christin \cite{christin_algorithms_2017} also uses ``decoupling'' to discuss algorithms in an organizational context, in an allied but somewhat different use. Christin draws on classic sociology of organizations work \cite{meyer_institutionalized_1977} to describe decoupling as when gaps emerge between how algorithms are used in practice by workers and how the organization's leadership assumes they are used, like when workers ignore or subvert algorithmic systems imposed on them by management. We could see such cases as unintentionally decoupled ML socio-technical systems in our use of the term, whereas ORES is intentionally decoupled.  There is also some use of ``decoupling'' in ML to refer to training separate models for members of different protected classes~\cite{dwork2018decoupled}, which is not the sense we mean, although this is in same spirit of going beyond a `one model to rule them all' strategy. 

As discussed in section~\ref{sec:collaborative_model_design}, we collaboratively consult and work with Wikipedians to design models that represent their ``emic'' concepts, which have meaning and significance in their communities. Our team works more as stewards of ML for a community, making decisions to support wikis based on our best judgment. In one sense, this can be considered a more coupled approach to a socio-technical ML system, as the people who legally own the servers do have unitary control over what models can be developed --- even if our team has a commitment to exercise that control in a more participatory way. Yet because of this commitment, the earlier stages of the ML process are still more decoupled than in most content moderation models for commercial social media platforms, where corporate managers hire contractors to label data using detailed instructions~\cite{gray2019ghost,roberts2019behind}.

While ORES's patterns around the tasks of model training, development, and deployment are somewhat-but-not-fully decoupled, tasks around model selection, use, and appropriation are highly decoupled. Given the open API, there is no barrier where we can selectively decide who can request a score from a classifier. We could have decided to make this part of ORES more strongly coupled, requiring users apply for revocable API keys to get ORES scores. This has important implications for how the models are regulated, limiting the direct power our team has over how models are used.  This requires that Wikipedians and volunteer tool developers govern and regulate the use of these ML models using their own structures, strategies, values, and processes.

\subsubsection{Decoupled model appropriation}
\label{sec:decoupled_model_appropriation}
ORES's open API sees a wide variety of uses.  The majority of the use is by volunteer tool developers who incorporate ORES's predictions into the user experiences they target with their user interfaces (e.g. a score in an edit review tool) and into decisions made by fully-automated bots (e.g. all edits above a threshold are auto-reverted). We have also seen appropriation by professional product teams at the Wikimedia Foundation (e.g. see section~\ref{sec:recent_changes_filters}) and other researchers (e.g., see section~\ref{sec:anonymous_and_tor_users}).  In these cases, we play at most a supporting role helping developers and researchers understand how to use ORES, but we do not direct these uses beyond explaining the API and how to use various features of ORES. 

This is relatively unusual for a Wikimedia Foundation-supported or even a volunteer-led engineering project, much less a traditional corporate product team. Often, products are deployed as changes and extensions of user-facing functionality on the site.  Historically, there have been disagreements between the Wikimedia Foundation and a Wikipedia community about what kind of changes are welcome (e.g., the deployment of ``Media Viewer" resulted in a standoff between the German Wikipedia community and the Wikimedia Foundation\footnote{\url{https://meta.wikimedia.org/wiki/Superprotect}}\cite{neotarf2014media}).  This tension between the values of the owners of the platform and the community of volunteers exists at least in part due to how software changes are deployed.  Indeed, we might have implemented ORES as a system that connected models to the specific user experiences that we valued.  Under this pattern, we might design UIs for patrolling or task routing based on our values --- as we did with an ML-assisted tool to support mentoring ~\cite{halfaker2014snuggle}. 

ORES represents a shift towards a far more decoupled approach to ML, which gives more agency to Wikipedian volunteers, in line with Wikipedia's long history more decentralized governance structures~\cite{forte2009decentralization}. For example, English Wikipedia and others have a Bot Approvals Group (BAG) \cite{halfaker2012bots, geiger2011lives} regulating which bots are allowed to edit articles in specific ways. This has generally been effective in ensuring that bot developers do not get into extended conflicts with the rest of the community or each other.~\cite{geiger2017operationalizing}.  If a bot were programmed to act against the consensus of the community --- with or without using ORES --- the BAG could shut the bot down without our help.

Beyond governance, the more decoupled and participatory approach to ML that ORES affords allows innovation by people who hold different values, points of view, and ideas than we do. For example, student contributors to Wikipedia were our main focus, but Ross was able to appropriate those predictions to support the student contributors he valued and whose concerns were important (see section~\ref{sec:ross_recommendations}).  Similarly, Tor users were not our main focus, but Tran et al. explored the quality of Tor user contributions to argue for for their high quality~\cite{tran2019tor}. 
}

\subsubsection{Demonstrative case: PatruBOT and Spanish Wikipedia}
\label{sec:patrubot_and_spanish_wikipedia}
Soon after we released a ``damaging" edit model for Spanish Wikipedia, a volunteer developer designed PatruBOT, a bot that automatically reverted any new edit to Spanish Wikipedia where ORES returned a score above a certain threshold. Our discussion spaces were soon bombarded with confused Spanish-speaking editors asking us why ORES did not like their edits.  We struggled to understand the complaints until someone told us about PatruBOT and showed us where its behavior was being discussed on Spanish Wikipedia. 

After inspecting the bot's code, we learned this case was about tradeoffs between precision/recall and false positives/negatives --- a common ML issue. PatruBOT's threshold for reverting was far too sensitive. ORES reports a prediction and a probability of confidence, but it is up to the local developers to decide if the bot will auto-revert edits classified as damage with a .90, .95, .99, or higher confidence. Higher thresholds minimize the chance a good edit will be mistakenly auto-reverted (false-positive), but also increase the chance that a bad edit will not be auto-reverted (false-negative).  Ultimately, we hold that each volunteer community should decide where to draw the line between false positives and false negatives, but we would could help inform these decisions.

{\color{\highlightcolor}
The Spanish Wikipedians held a discussion about PatruBOT's many false-positives.\footnote{\url{https://es.wikipedia.org/wiki/Wikipedia:Caf\%C3\%A9/Portal/Archivo/Miscel\%C3\%A1nea/2018/04\#Parada_de_PatruBOT}}  Using wiki pages, they crowdsourced an audit of PatruBOT's behavior.\footnote{\url{https://es.wikipedia.org/wiki/Wikipedia:Mantenimiento/Revisi\%C3\%B3n_de_errores_de_PatruBOT\%2FAn\%C3\%A1lisis}}  They came to a consensus that PatruBOT was making too many mistakes and it should stop until it could be fixed. A volunteer administrator was quickly able to stop PatruBOTs activities by blocking its user account, which is a common way bots are regulated. This was entirely a community governed activity that  required no intervention of our team or the Wikimedia Foundation staff.

This case shows how Wikipedian stakeholders do not need to have an advanced understanding in ML evaluation to meaningfully participate in a sophisticated discussion about how, when, why, and under what conditions such classifiers should be used. Because of the API-based design of the ORES system, no actions are needed on our end once they make a decision, as the fully-automated bot is developed and governed by Spanish Wikipedians and their processes.  In fact, upon review of this case, we were pleased to see that SeroBOT took the place of PatruBOT in 2018 and continues to auto-revert vandalism today --- albeit with a higher confidence threshold.

\subsubsection{Stewarding model design/deployment}
\label{sec:stewarding_model_design}
As maintainers of the ORES servers, our team does retain primary jurisdiction over every model: we must approve each request and can change or remove models as we see fit. We have rejected some requests that community members have made, specifically rejecting multiple requests for an automated plagiarism detector. However, we rejected those requests more because of technical infeasibility, rather than principled objections. This would require a complete-enough database of existing copyrighted material to compare against, as well as immense computational resources that other content scoring models do not require.

In the future, it is possible that our team will have to make difficult and controversial decisions about what models to build. First, demand for new models could grow such that the team could not feasibly accept every request, because of a lack of human labor and/or computational resources. We would have to implement a strategy for prioritizing requests, which like all systems for allocating scarce resources, would have embedded values that benefit some more than others. Our team has been quite small: we have never had more than 3 paid staff and 3 volunteers at any time, and no more than 2 requests typically in progress simultaneously. This may become a different enterprise if it grew to dozens or hundreds of people --- as is the size of ML teams at major social media platforms --- or received hundreds of requests for new models a day. 

Our team could also receive requests for models in fundamental contradiction with our values: for example, classifying demographic profiles of an anonymous editor from their editing behavior, which would raise privacy issues and could be used in discriminatory ways. We have not formally specified our policies on these kinds of issues, which we leave to future work. Wikipedia's own governance systems started out informal and slowly began to formalize as was needed~\cite{forte2009decentralization}, with periodic controversies prompting new rules and governing procedures. Similarly, we expect that as issues like these arise, we will respond to them iteratively, building to when formally specified policies and structures will be necessary. However, we note `just-in-time' approaches to governance have led to user-generated content platforms deferring decisions on critical issues \cite{vaidhyanathan2018antisocial}.

ORES as a technical and socio-technical system can be more or less tightly coupled onto existing social systems, which can dramatically change their operation and governance. For example, the team could decide that any new classifier for a local language wiki would only be deployed if the entire community held a discussion and reached a consensus in favor of it. Alternatively, the team could decide that any request would be accepted and implemented by default, but if a local language wiki reached a consensus against a particular classifier, they would take it down. These hypothetical decisions by our team would implement two rather different governance models around ML, which would both differ from the current way the team operates.

\subsubsection{Demonstrative case: Article quality modeling for Dutch Wikipedia}
\label{sec:dutch_wikipedia_and_article_quality}
After meeting our team at an in-person Wikimedia outreach event, two Dutch Wikipedians (Ciell and RonnieV) reached out to us to build an article quality model for Dutch Wikipedia.  After working with them to understand how Dutch Wikipedians thought about article quality and what kind of digital traces we might be able to extract, Ciell felt it was important to reach out to the Dutch community to discuss the project. She made a posting on \emph{De kroeg}\footnote{\url{https://nl.wikipedia.org/wiki/Wikipedia:De_kroeg/Archief/20200529\#Wikimedia_Hackathon_2020}} --- a central local discussion space.

At first, the community response to the idea of an article quality model was very negative, on the grounds that an algorithm could not measure the aspects of quality they cared about. They also noted that while ORES may work well for English Wikipedia, this model could not be used on Dutch Wikipedia due to differences in language and quality standards. After discussing that the model could be designed to work with training data from their wiki and using their specific quality scale, there was some support. However, the crucial third point was that adding a model to the ORES service would not necessarily change the user interface for everyone, if the community did not want to use it in that way. It would be a service that anyone could query on their own, which could be implemented as \emph{opt-in} feature via bespoke code, which in this case was a Javascript gadget that the community could manage --- as it currently manages many other opt-in add-on extensions and gadgets. The tone of the conversation immediately changed, and the idea of running a trial to explore the usefulness of ORES article quality predictions was approved.  

This case illustrates two key issues. First, the Dutch Wikipedians were hesitant to approve a new ML-based feature that would affect all users' experience of the entire wiki, but they were much more interested in an optional opt-in ML feature. Second, they rejected adopting models originally tailored for English Wikipedia, explicitly discussing how they conceptualized quality differently. As ORES operates as a service they can use on their own terms, they can direct the design and use of the models. This would have been far more controversial if a ML-based article quality tool had been designed in more standard coupled product engineering model, like if we added a feature that surfaced article quality predictions to everyone. 
}

\section{Discussion and future work}
\label{sec:discussion_and_future_work}
{\color{\highlightcolor}
\subsection{Participatory machine learning}
In a world increasingly dominated by for-profit user-generated content platforms --- often marketed by their corporate owners as ``communities''~\cite{gillespie2018custodians} --- Wikipedia is an anomaly. While the non-profit Wikimedia Foundation has only a fraction of the resources as Facebook or Google, the unique principles and practices in the broad Wikipedia/Wikimedia movement are a generative constraint. ORES emerged out of this context, operating at the intersection of a pressing need to deploy efficient machine learning at scale for content moderation, but to do so in ways that enable volunteers to develop and deploy advanced technologies on their own terms. Our approach is in stark contrast to the norm in machine learning research and practice, which involves a more tightly-coupled, top-down mode of developing the most precise classifiers for a known ground truth, then wrapping those classifiers in a complete technology for end-users, who must treat them as black boxes.

The more wiki-inspired approach to what we call ``participatory machine learning'' imagines classifiers to be just as provisional and open to criticism, revision, and skeptical reinterpretation as the content of Wikipedia's encyclopedia articles. And like Wikipedia articles, we suspect some classifiers will be far better than others based on how volunteers develop and curate them, for various definitions of ``better'' that are already being actively debated. Our demonstrative cases and exploratory work by Smith et al. based on ORES~\cite{smith2020keeping} briefly indicate how volunteers have collectively engaged in sophisticated discussions about how they ought to use machine learning. ORES's fully open, reproducible/auditable code and data pipeline --- from training data to models and  scored predictions --- enables a wide range of new collaborative practices. ORES is a more socio-technical and CSCW-oriented approach to issues in the FAccTML space, where attention is often placed on mathematical and technical solutions, like interactive visualizations for model interpretability or formal guarantees of operationalized definitions of fairness~\cite{selbst_fairness_2019,mulligan2019}.

ORES also represents an innovation in openness in that it decouples several activities that have typically all been performed by managers/engineers or those under their direct supervision and control: deciding what will be modeled, labeling training data, choosing or curating training data, engineering features, building models to serve predictions, auditing predictions for false positives/negatives, and developing interfaces or automated agents that act on those predictions.  As our cases have shown, people with extensive contextual and domain expertise in an area can make well-informed decisions about curating training data, identifying false positives/negatives, setting thresholds, and designing interfaces that use scores from a classifier. In decoupling these actions, ORES helps delegate these responsibilities more broadly, opening up the structure of the socio-technical system and expanding who can participate in it. In the next section, we introduce this concept of decoupling more formally through a comparison to a quite different ML system, then in later sections show how ORES has been designed to these ends.

\subsection{What is decoupling in machine learning?}
A 2016 ProPublica investigation~\cite{angwin2016machine} raised serious allegations of racial biases in a ML-based tool sold to criminal courts across the US. The COMPAS system by Northpointe, Inc. produced risk scores for defendants charged with a crime, to be used to assist judges in determining if defendants should be released on bail or held in jail until their trial. This expos\'e began a wave of academic research, legal challenges, journalism, and organizing about a range of similar commercial software tools that have saturated the criminal justice system. Academic debates followed over what it meant for such a system to be ``fair'' or ``biased''~\cite{adler_auditing_2018,berk_fairness_2018,dressel_accuracy_2018,zafar_fairness_2017,corbett-davies_algorithmic_2017} As Mulligan et al.~\cite{mulligan2019} discuss, debates over these ``essentially contested concepts'' often focused on competing mathematically-defined criteria, like equality of false positives between groups, etc.

When we examine COMPAS, we must admit that we feel an uneasy comparison between how it operates and how ORES is used for content moderation in Wikipedia. Of course, decisions about what is kept or removed from Wikipedia are of a different kind of social consequence than decisions about who is jailed by the state. However, just as ORES gives Wikipedia's human patrollers a score intended to influence their gatekeeping decisions, so does COMPAS give judges a similarly-functioning score. Both are trained on data that assumes a knowable ground truth for the question to be answered by the classifier. Often this data is taken from prior decisions, heavily relying on found traces produced by a multitude of different individuals, who brought quite different assumptions and frameworks to bear when originally making those decisions.

Yet comparing the COMPAS suite with ORES as socio-technical systems, one of the more striking differences to us --- beyond inherent issues in using ML in criminal justice systems, see \cite{barabas_abolish_2020} for a review --- is how tightly coupled COMPAS is. This is less-often discussed, particularly as a kind of meta-value embeddable in design. COMPAS is a standalone turnkey vendor solution, developed end-to-end by Northpointe, based on their particular top-down values. The system is a deployable set of pretrained models trained on data from nationwide ``norm groups''  that can be dropped into any criminal justice system's IT stack. If a client wants retraining of the ``norm groups'' (or other modifications) this additional service must be purchased from Northpointe.\footnote{\url{https://web.archive.org/web/20200417221925/http://www.northpointeinc.com/files/downloads/FAQ_Document.pdf}}

COMPAS scores are also only to be accessed through an interface provided by Northpointe.\footnote{\url{https://web.archive.org/web/20200219210636/http://www.northpointeinc.com/files/downloads/Northpointe_Suite.pdf}} This interface shows such scores as part of workflows within the company's flagship product Northpointe Suite --- a enterprise management system criminal justice organizations. Many debates about COMPAS and related systems have placed less attention on the models as an element of a broader socio-technical software system within organizations, although there are notable exceptions like Christin's ethnographic work on algorithmic decision making in criminal justice, who raises similar organizational issues~\cite{christin_algorithms_2017,cristin2018predictive}. Several of our critiques of tightly coupled ML systems are also similar to the `traps' of abstraction that Selbst et al.~\cite{selbst_fairness_2019} see ML engineers fall into.

Given such issues, many have called for public policy solutions, such as requiring public sector ML systems to have source code and training data released, or at least a structured transparency report (e.g. \cite{gebru2018datasheets,mitchell2019model}). We agree with this, but it is not simply that ORES is open source and open data, while COMPAS is not. It is just as important to have an open socio-technical system that flexibly accommodates the kinds of decoupled decision-making around data, model building and tuning, and decisions about how scores will be presented and used. A more decoupled approach does not mitigate potential harms, but can provide better entrypoints for auditing and criticism. 

Extensive literature has discussed problems in taking found trace data from prior decisions as ground truth, particularly in institutions with histories of systemic bias \cite{boyd2011six,gitelman2013raw}. This has long been our concern with the standard approach in ML for content moderation in Wikipedia, which often uses the entire set of past edit revert decisions as ground truth. These concerns are rising around the adoption of ML in other domains where institutions may have an easily-parsable dataset of past decisions, from social services to hiring to finance~\cite{eubanks2018automating}. However, we see a gap in the literature for a uniquely CSCW-oriented view of ML systems as software-enabled organizations --- with resonances to classic work on Enterprise Resource Planning systems~\cite[e.g.][]{pollock2008software}. In taking this broader view, argue that instead of attempting to de-bias problematic datasets for a single model, we seek to broaden participation and offer multiple contradictory models trained on different datasets.



In a highly coupled socio-technical machine learning system like COMPAS, a single set of people are responsible for all aspects of the ML process: labeling training data, engineering features, building models with various algorithms and parameterizations, choosing between these models, scoring items using models, and building interfaces or agents using those scores. As COMPAS is designed (and marketed), there is little capacity for people outside Northpointe, Inc. to intervene at any stage. Nor is there much of a built-in capacity to scaffold on new elements to this workflow at various points, such as introducing auditing or competing models at any point. This is a common theme in the auditing literature in and around FAccT~\cite{sandvig2014auditing}, which often emphasizes ways to reverse engineer how models operate. In the next section, we discuss how ORES is more coupled in some aspects, while less coupled in others, which draws the demonstrative cases to describe the architectural and organizational decisions that produced such a configuration.

\subsection{Cases of decoupling and appropriation of machine learning models}
\leadin{Multiple independent classifiers trained on multiple independent data sets}
Many ML-based workflows presume a single canonical training data set approximating a ground truth, producing a model to be deployed at scale. ORES as socio-technical system is designed to support many co-existing and even contradictory training data sets and classifiers. This was initially a priority because of the hundreds of different wiki projects across languages that are edited independently. Yet this design constraint also pushed us to develop a system where multiple independent classifiers can be trained on different datasets within the same wiki project, based on different ideas about what `quality' or `damage' are.

Having independent co-existing classifiers is a key sign of a more decoupled system, as different people and publics can be involved in the labeling and curation of training data and in model design and auditing (which we discuss later). This decoupling is supported both in the architectural decisions outlined in our sections on scaling and API design, as well as in how our team has somewhat formalized processes for wiki contributors to request their own models. As we discussed, this is not a fully decoupled approach were anyone can build their own model with a few clicks, as our team still plays a role in responding to requests, but it is a far more decoupled approach to training data than in prior uses of ML in Wikipedia and in many other real-world ML systems.

\leadin{Supporting latent user-needs}
While the ORES socio-technical system is implemented in a somewhat-but-not-fully decoupled mode for model design and iteration, appropriation of models by developers and researchers is highly decoupled.  Decoupling the maintenance of a model and the adoption of the model seems to have lead to adoption of ORES as a commons commodity --- a reusable raw component that can be incorporated into specific products and that is owned and regulated as a common resource\cite{ostrom1990governing}.  The ``damage" detection models in ORES allow \emph{damage detection} or the reverse (\emph{good edit detection}) to be much more easily appropriated into end-user products and analyses.  In section~\ref{sec:anonymous_and_tor_users}, we showed how related research did this for exploring Tor users' edits.  We also showed in section~\ref{sec:ross_recommendations} how Ross re-imagined the article quality models as a strategy for building work suggestions.  We had no idea that this model would be useful for this, and the developers and researchers did not need to ask us for permission to use ORES this way.  

When we chose to provide an open API---to decouple model adoption---we stopped designing only for cases we imagined, moving to support latent user-needs~\cite{meyera2001perspective}.  By providing infrastructure for others to use as they saw fit, we are enabling needs that we cannot anticipate.  This is a less common approach in ML, where it is more common for a team to design a specific ML-supported interface for users end-to-end.  Indeed, this pattern was the state of ML in Wikipedia before ORES, and as we argue in section~\ref{sec:machine_learning_resource_distribution}, this lead to many users' needs being unaddressed, with substantial systemic consequences. 

\leadin{Enabling community governance authority}
Those who develop and maintain technical infrastructure for software-mediated social systems often gain an \textit{incidental jurisdiction} over the entire social system, able to make technical decisions with major social impacts. \cite{lessig1999code}  In the case of Wikipedia, decoupling the maintenance of ML resources from their application pushes back against this pattern. ORES has enabled the communities of Wikipedia to more easily make governance decisions around ML by not requiring them to negotiate as much with us, the model maintainers.  In section~\ref{sec:patrubot_and_spanish_wikipedia}, we discuss how Spanish Wikipedians decided to approve PatruBOT to run in their wiki, then shut down the bot again without petitioning us to do so.  Similarly, in section~\ref{sec:dutch_wikipedia_and_article_quality}, we discuss how Dutch Wikipedians were much more interested in experimenting with ORES when they learned how they would maintain control of their user experience. 

\leadin{The powerful role of auditing in decoupled ML systems}
A final example of the importance of decoupled ML systems is in auditing. In our work with ORES, we recruit models' users to perform an audit after every deployment or substantial model change.  We have found misclassification reports to be an effective boundary object~\cite{leigh2010not} for communicating about the meaning a model captures and does not capture.  In section~\ref{sec:italian_wikipedia_thematic_analysis}, we showed how a community of non-subject matter expert volunteers helped us identify a specific issue related to a bug in our modeling pipeline via an audit and thematic analysis. Wikipedians imagined the behavior they desired from a model and contrasted that to actual behavior, and in response, our engineering team imagined the technical mechanisms for making ORES behavior align with their expectations.  

Wikipedians were also able to use auditing to make value-driven governance decisions.  In section~\ref{sec:patrubot_and_spanish_wikipedia}, we showed evidence of critical reflection on the current processes and the role of algorithms in quality control processes.  While the participants were discussing issues that ML experts would refer to as \emph{model fitness} or \emph{precision}, they did not use that language or a formal understanding of model fitness.  Yet they were able to effectively determine both the fitness of PatruBOT and make decisions about what criteria were important to allow the continued functioning of the bot.  They did this using their own notions of how the bot should and should not behave and by looking at specific examples of the bots behavior in context.

Eliciting this type of critical reflection and empowering users to engage in their own choices about the roles of algorithmic systems in their social spaces has typically been more of a focus from the Critical Algorithms Studies literature, which comes from a more humanistic and interpretivist social science perspective (e.g.~\cite{barocas2013governing, kitchin2017thinking}).  This literature also emphasizes a need to see algorithmic systems as dynamic and constantly under revision by developers \cite{seaver2017algorithms} --- work that is invisible in most platforms, but is intentionally foregrounded in ORES.  We see great potential for building new systems for supporting the crowd-sourced collection and interpretation of this type of auditing data.  Future work should explore strategies for supporting auditing as a normal part of model design, deployment, and maintenance.
}

\subsection{Design implications}
In many user-generated content platforms, the technologies that mediate social spaces are controlled top-down by a single organization. Many social media users express frustration over their lack of agency around various understandings of ``the algorithm.'' \cite{burrell2019}. We have shown how if such an organization seeks to involve its users and stakeholders more around ML, they can employ a \emph{decoupling} strategy like that of ORES, where professionals with infrastructural expertise build and serve ML models at scale, while other stakeholders curate training data, audit model performance, and decide where and how the ML models will be used. The demonstrative cases show the feasibility and benefits of decoupling the ML modeling service from the curation of training data and the implementation of ML scores in interfaces and tools, as well as in moving away from a single ``one classifier to rule them all'' and towards giving users agency to train and serve models on their own. 

With such a service, a wide range of people play critical roles in the governance of ML in Wikipedia, which go beyond what they would be capable of if ORES were simply another ML classifier hidden behind a single-purpose UI --- albeit with open-source code and training data, as prior ML classifiers in Wikipedia were. Since ORES has been in service, more than twenty-five times more ML models have been built and served at Wikipedia's scale than in the entire 15 year history of Wikipedia prior. Open sourcing the data and model code behind these original pre-ORES models did not lead to a proliferation of alternatives, while ORES as a socio-infrastructural service did. This is because ORES as a technical and socio-technical system reduces ``incidental complexities'' \cite{mays1994forging} involved in developing the systems necessary for deploying ML in production and at scale.

{\color{\highlightcolor}
Our design implications for organizations and platforms is to take this literally: to run open ML as a service so that users can build their own models with training datasets they provide, which serve predictions using open APIs, and support activities like \emph{dependency injection} and \emph{threshold optimization} for auditing and re-appropriation. Together with a common set of discussion spaces like the ones Wikipedians used, these can enable the re-use of models by a broader audience and make space for reflective practices such as model auditing, decision-making about thresholds, or the choice between different classifiers trained on different training datasets.  People with varying backgrounds and expertise in programming and ML (at least in our field site) have an interest in participating in such governance activities and can effectively coordinate common understandings of what a ML model is doing and whether or not that is acceptable to them.

\subsection{Limitations and future work}
Observing ORES in practice suggests avenues of future work toward crowd-based auditing tools.  As our cases demonstrate, auditing of ORES' predictions and mistakes has become a popular activity both during quality assurance checks after deploying a model (see section~\ref{sec:italian_wikipedia_thematic_analysis}) and during community discussions about how a model should be used (see section~\ref{sec:patrubot_and_spanish_wikipedia}).  Even though we did not design interfaces for discussion and auditing, some Wikipedians have used unintended affordances of wiki pages and MediaWiki's template system to organize processes for flagging false positives and calling attention to them.  This process has proved invaluable for improving model fitness and addressing critical issues of bias against disempowered contributors (see section~\ref{sec:anonymous_and_tor_users}).  

To better facilitate this process, future system builders should implement structured means to refute, support, discuss, and critique the predictions of machine models.  With a structured way to report what machine prediction gets right and wrong, the process of reporting mistakes could be streamlined---making the negotiation of meaning in a machine learning model part of its every day use.  This could also make it easier to perform the thematic audits we saw in section~\ref{sec:italian_wikipedia_thematic_analysis}.  For example, a structured database of ORES mistakes could be queried in order to discover groups of misclassifications with a common theme.  By supporting such an activity, we are working to transfer more power from ourselves (the system owners) and to our users.  Should one of our models develop a nasty bias, our users will be more empowered to coordinate with each other, show that the bias exists and where it causes problems, and either get the modeling pipeline fixed or even shut down a problematic usage pattern---as Spanish Wikipedians did with PatruBOT.

ORES has become a platform for supporting researchers, who use ORES both in support of and in comparison to their own analytical or modeling work.  For example, Smith et al. used ORES as a focal point of discussion about values applied to machine learning models and their use~\cite{smith2020keeping}.  ORES is useful for this because it is not only an interesting example for exploration, but also because it has been incorporated as part of the essential infrastructure of Wikipedia~\cite{vaseva2019you}.   Dang et al.~\cite{dang2016quality,dang2017end} and Joshi et al.~\cite{joshi2020detecting} use ORES as a baseline from which to compare their novel modeling work.  Halfaker et al.~\cite{halfaker2017interpolating} and Tran et al.~\cite{tran2019tor} use ORES scores directly as part of their analytical strategies.  And even Yang et al. used \emph{Wiki labels} and some of our feature extraction systems to build their own models~\cite{yang2017identifying}.  We look forward to what socio-technical explorations, model comparisons, and analytical strategies will be built on this research platform by future work.

We also look forward to what those from the fields around Fairness, Accountability, and Transparency in ML and Critical Algorithm Studies can ask, do, and question about ORES. Most of the studies and critiques of \emph{subjective algorithms}~\cite{tufekci2015algorithms} focus on for-profit or governmental organizations that are resistant to external interrogation. Wikipedia is one of the largest and arguably more influential information resources in the world, and decisions about what is and is not represented have impacts across all sectors of society.  The algorithms that ORES makes available are part of the decision process that leads to some people's contributions remaining and others being removed.  This is a context where algorithms have massive social consequence, and we are openly exploring transparent and open processes to help address potential issues and harms.

There is a large body of work exploring how biases manifest and how unfairness can play out in algorithmically mediated social contexts. ORES would be an excellent place to expand the literature within a specific and important field site.  Notably, Smith et al. have used ORES as a focus for studying \emph{Value-Sensitive Algorithm Design} and highlighting convergent and conflicting values~\cite{smith2020keeping}.  We see great potential for research exploring strategies for more effectively encoding these values in both ML models and the tools/processes that use them on top of open machine learning services like ORES. We are also very open to the likelihood that our more decoupled approaches could still be reinforcing some structural inequalities, with ORES solving some issues but raising new issues. In particular, the decentralization and delegation to a different set of self-selected Wikipedians on local language versions may raise new issues, as past work has explored the hostility and harassment that can be common in some Wikipedia communities \cite{menking2015heart}. Any approach that seeks to expand access to a potentially harmful technology should also be mindful about unintended uses and abuses, as adversaries and harms in the content moderation space are numerous \cite{neff2016automation,massanari2017gamergate,wolf2017we}.

Finally, we also see potential in allowing Wikipedians to freely train, test, and use their own prediction models without our engineering team involved in the process.  Currently, ORES is only suited to deploy models that are trained and tested by someone with a strong modeling and programming background, and we currently do that work for those who come to us with a training dataset and ideas about what kind of classifier they want to build.  That does not necessarily need to be the case.  We have been experimenting with demonstrating ORES model building processes using Jupyter Notebooks~\cite{kluyver2016jupyter} \footnote{e.g. \url{ https://github.com/wiki-ai/editquality/blob/master/ipython/reverted_detection_demo.ipynb}} and have found that new programmers can understand the work involved.  This is still not the fully-realized accessible approach to crowd-developed machine prediction, where all of the incidental complexities involved in programming are removed from the process of model development and evaluation.  Future work exploring strategies for allowing end-users to build models that are deployed by ORES would surface the relevant HCI issues involved and the changes to the technological conversations that such a margin-opening intervention might provide, as well as be mindful of potential abuses and new governance issues.
}

\section{Conclusion}
\label{sec:conclusion}
{\color{\highlightcolor}
In this `socio-technical systems paper,' we first discussed ORES as a technical system: an open API for providing access to machine learning models for Wikipedians.  We then discussed the socio-technical system we have developed around ORES that allows us to encode communities emic concepts in their models, collaboratively and iteratively audit their performance, and support broad appropriation of the models both within Wikipedia's editing community and in the broader research community.  We have also shown a series of \emph{demonstrative cases} how these concepts are negotiated, audits are performed, and appropriation has taken place.  This system, the observations, and the cases show a deep view of a technical system and the social structures around it. In particular, we analyze this arrangement as a more \textit{decoupled} approach to machine learning in organizations, which we see as a more CSCW-inspired approach to many issues being raised around the fairness, accountability, and transparency of machine learning.
}

\section{Acknowledgements}
\label{sec:acknowledgements}
This work was funded in part by the Gordon \& Betty Moore Foundation (Grant GBMF3834) and Alfred P. Sloan Foundation (Grant 2013-10-27), as part of the Moore-Sloan Data Science Environments grant to UC-Berkeley, and directly by the Wikimedia Foundation.  Thanks to our Wikipedian collaborators for sharing their insights and reviewing the descriptions of our collaborations: Chaitanya Mittal (chtnnh), Helder Geovane (He7d3r), Goncalo Themudo (GoEThe), Ciell, Ronnie Velgersdijk (RonnieV), and Rotpunkt.  


\bibliographystyle{ACM-Reference-Format}
\bibliography{refs}

\pagebreak
\appendix
\section{Appendix}
\label{sec:appendix}
\subsection{Empirical access patterns}
\label{sec:appendix.empirical_access_patterns}
\begin{figure}[h]
\centering
\begin{subfigure}[t]{\columnwidth}
  \centering
  \includegraphics[width=.75\textwidth]{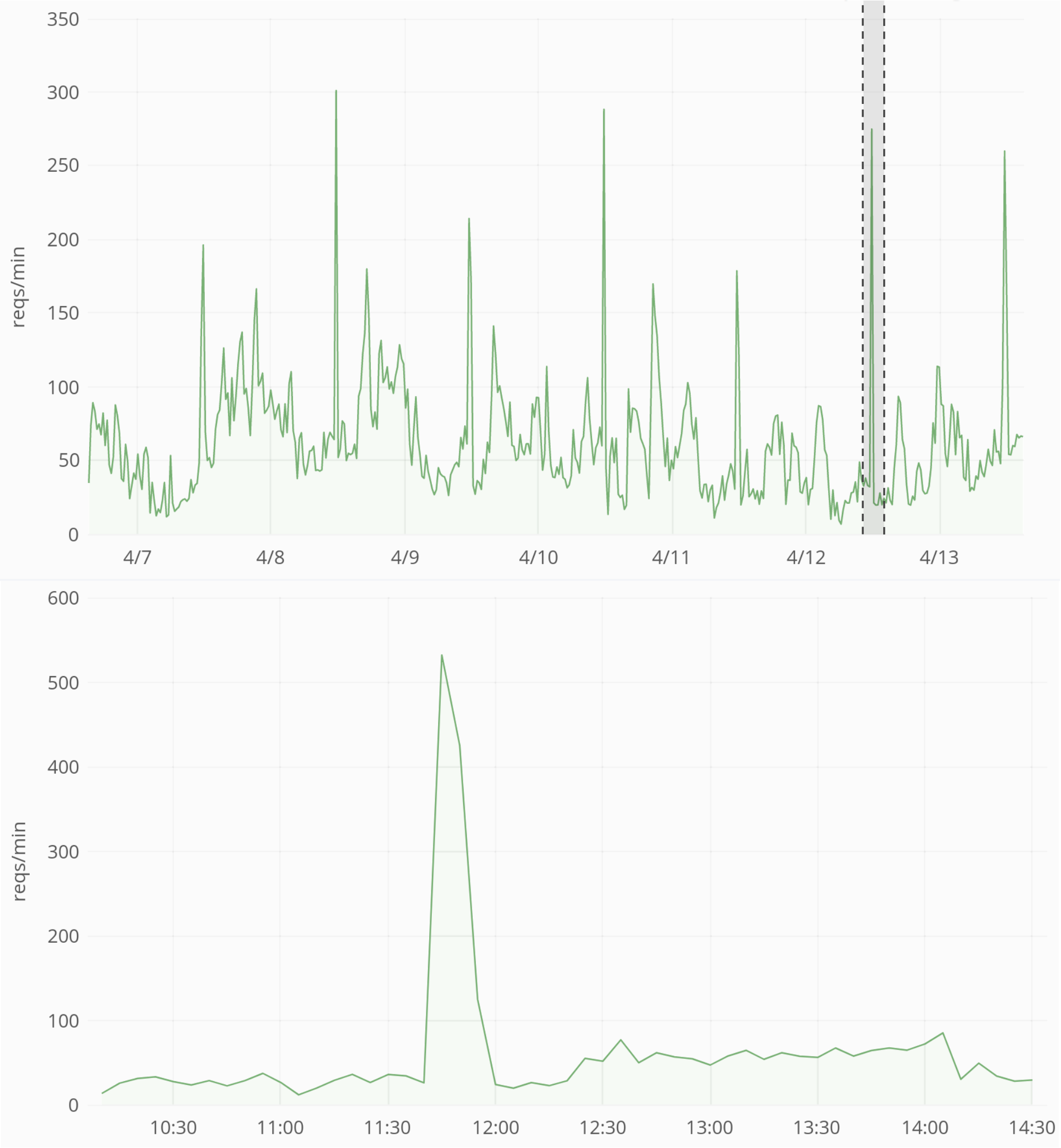}
  \caption{External requests per minute with a 4 hour block broken out to highlight a sudden burst of requests}
  \label{fig:ores_request_rate}
\end{subfigure}\\
\begin{subfigure}[t]{\columnwidth}
  \centering
  \includegraphics[width=.75\textwidth]{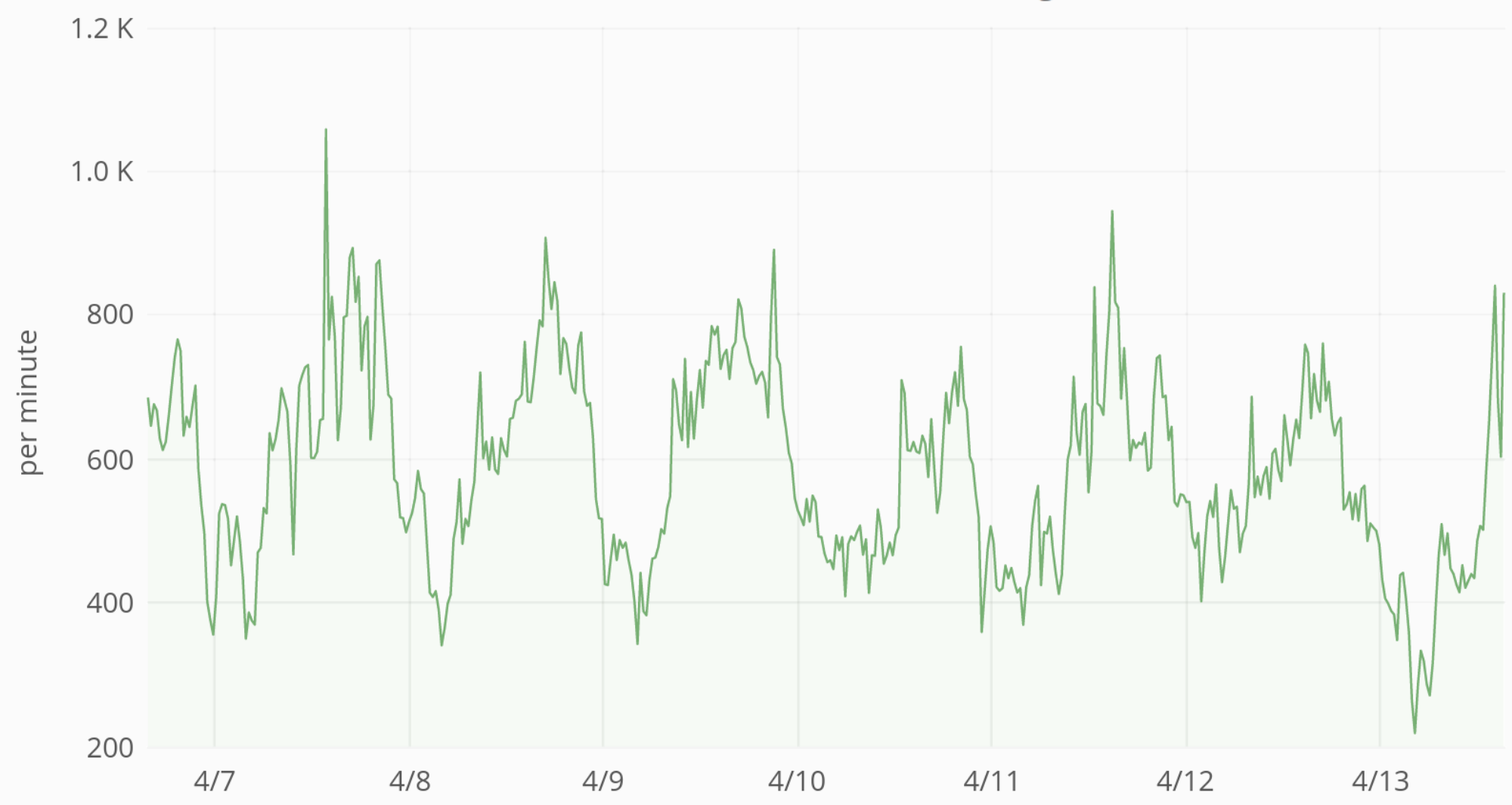}
  \caption{Precaching requests per minute}
  \label{fig:ores_precache_rate}
\end{subfigure}
\caption{Request rates to the ORES service for the week ending on April 13th, 2018}
\label{fig:ores_activity}
\end{figure}

The ORES service has been online since July 2015\cite{halfaker2015artificial}.  Since then, usage has steadily risen as we've developed and deployed new models and additional integrations are made by tool developers and researchers.  Currently, ORES supports 78 different models and 37 different language-specific wikis.

Generally, we see 50 to 125 requests per minute from external tools that are using ORES' predictions (excluding the MediaWiki extension that is more difficult to track).  Sometimes these external requests will burst up to 400-500 requests per second.  Figure~\ref{fig:ores_request_rate} shows the periodic and ``bursty'' nature of scoring requests received by the ORES service.  For example, every day at about 11:40 UTC, the request rate jumps---most likely a batch scoring job such as a bot.

Figure~\ref{fig:ores_precache_rate} shows the rate of precaching requests coming from our own systems.  This graph roughly reflects the rate of edits that are happening to all of the wikis that we support since we'll start a scoring job for nearly every edit as it happens.  Note that the number of precaching requests is about an order of magnitude higher than our known external score request rate.  This is expected, since Wikipedia editors and the tools they use will not request a score for every single revision.  This is a computational price we pay to attain a high cache hit rate and to ensure that our users get the quickest possible response for the scores that they \emph{do} need.

Taken together these strategies allow us to optimize the real-time quality control workflows and batch processing jobs of Wikipedians and their tools.  Without serious effort to make sure that ORES is practically fast and highly available to real-time use cases, ORES would become irrelevant to the target audience and thus irrelevant as a boundary-lowering intervention.  By engineering a system that conforms to the work-process needs of Wikipedians and their tools, we've built a systems intervention that has the potential gain wide adoption in Wikipedia's technical ecology.

\subsection{Explicit pipelines}
\label{sec:appendix.explicit_pipelines}
Within ORES system, each group of similar models have explicit model training pipelines defined in a repo.  Currently, we support 4 general classes of models:
\begin{itemize}
    \item \emph{editquality}\footnote{\url{https://github.com/wikimedia/editquality}} -- Models that predict whether an edit is ``damaging'' or whether it was save in ``good-faith''
    \item \emph{articlequality}\footnote{\url{https://github.com/wikimedia/articlequality}} -- Models that predict the quality of an article on a scale
    \item \emph{draftquality}\footnote{\url{https://github.com/wikimedia/draftquality}} -- Models that predict whether new articles are spam or vandalism
    \item \emph{drafttopic}\footnote{\url{https://github.com/wikimedia/drafttopic}} -- Models that predict the general topic space of articles and new article drafts
\end{itemize}

Within each of these model repositories is a collection of facility for making the modeling process explicit and replay-able. Consider the code shown in figure~\ref{fig:english_damaging_makefile} that represents a common pattern from our model-building Makefiles.

Essentially, this code helps someone determine where the labeled data comes from (manually labeled via the Wiki Labels system).  It makes it clear how features are extracted (using the \texttt{revscoring extract} utility and the \texttt{feature\_lists.enwiki.damaging} feature set).  Finally, this dataset of extracted features is used to cross-validate and train a model predicting the ``damaging'' label and a serialized version of that model is written to a file.  A user could clone this repository, install the set of requirements, and run \texttt{make enwiki\_models} and expect that all of the data-pipeline would be reproduced, and an equivalent model obtained.  

By explicitly using public resources and releasing our utilities and Makefile source code under an open license (MIT), we have essentially implemented a turn-key process for replicating our model building and evaluation pipeline.  A developer can review this pipeline for issues knowing that they are not missing a step of the process because all steps are captured in the Makefile.  They can also build on the process (e.g. add new features) incrementally and restart the pipeline.  In our own experience, this explicit pipeline is extremely useful for identifying the origin of our own model building bugs and for making incremental improvements to ORES' models.

At the very base of our Makefile, a user can run \texttt{make models} to rebuild all of the models of a certain type.  We regularly perform this process ourselves to ensure that the Makefile is an accurate representation of the data flow pipeline.  Performing complete rebuild is essential when a breaking change is made to one of our libraries or a major improvement is made to our feature extraction code.  The resulting serialized models are saved to the source code repository so that a developer can review the history of any specific model and even experiment with generating scores using old model versions.  This historical record of past models has already come in handy for audits of past model behavior. 

\begin{figure}[h]
        \makebox{\hrulefill}{
        \small
        \begin{verbatim}
datasets/enwiki.human_labeled_revisions.20k_2015.json:
    ./utility fetch_labels \
        https://labels.wmflabs.org/campaigns/enwiki/4/ > $@

datasets/enwiki.labeled_revisions.w_cache.20k_2015.json: \
        datasets/enwiki.labeled_revisions.20k_2015.json
    cat $< | \
        revscoring extract \
            editquality.feature_lists.enwiki.damaging \
            --host https://en.wikipedia.org \
            --extractor $(max_extractors) \
            --verbose > $@

models/enwiki.damaging.gradient_boosting.model: \
        datasets/enwiki.labeled_revisions.w_cache.20k_2015.json
    cat $^ | \
    revscoring cv_train \
        revscoring.scoring.models.GradientBoosting \
        editquality.feature_lists.enwiki.damaging \
        damaging \
        --version=$(damaging_major_minor).0 \
        -p 'learning_rate=0.01' \
        -p 'max_depth=7' \
        -p 'max_features="log2"' \
        -p 'n_estimators=700' \
        --label-weight $(damaging_weight) \
        --pop-rate "true=0.034163555464634586" \
        --pop-rate "false=0.9658364445353654" \
        --center --scale > $@
        \end{verbatim}
        \hrule
        \normalsize}
        \caption{Makefile rules for the English damage detection model from \url{https://github.com/wiki-ai/editquality}}
        \label{fig:english_damaging_makefile}
\end{figure}

\end{document}